\documentclass[apjl]{emulateapj}
\usepackage{comment}
\usepackage{ifthen}
\usepackage{amsmath}
\usepackage{amsfonts}
\usepackage{amssymb}
\usepackage{wasysym}
\usepackage{subfigure}

\newboolean{emulateapj}
\setboolean{emulateapj}{true}

\newboolean{astroph}
\setboolean{astroph}{true}
\shortauthors{Kipping et al.}
\shorttitle{I. The Hunt for Exomoons with Kepler (HEK)}
\ifthenelse{\boolean{emulateapj}}{
    
}{
    
}
\newcommand{\luna}{{\tt LUNA}}
\newcommand{\multi}{{\sc MultiNest}}
\newcommand{\ttvim}{{\tt ttvim.f}}

\begin{document}

%% Titlepage
\title {The Hunt for Exomoons with Kepler (HEK):\\
 I. Description of a New Observational Project}

%% Authors
\author{
	{\bf 
D.~M.~Kipping\altaffilmark{1,2}, 
G.~\'A.~Bakos\altaffilmark{3}, 
L.~Buchhave\altaffilmark{4}, 
D.~Nesvorn\'{y}\altaffilmark{5}, 
A.~Schmitt\altaffilmark{6}}
}
\altaffiltext{1}{Harvard-Smithsonian Center for Astrophysics,
	Cambridge, MA 02138, USA; email: dkipping@cfa.harvard.edu}

\altaffiltext{2}{NASA Carl Sagan Fellow}

\altaffiltext{3}{Dept. of Astrophysical Sciences, Princeton University,
Princeton, NJ 05844, USA}

\altaffiltext{4}{Niels Bohr Institute, Copenhagen University, Denmark}

\altaffiltext{5}{Dept. of Space Studies, Southwest Research Institute, 
1050 Walnut St., Suite 300, Boulder, CO 80302, USA}

\altaffiltext{6}{Citizen Science}

%% EOF authors

% #####################################################################
%% abstract
\begin{abstract}
%++++++++++++++++++++++++++++++++++++++++++++++++++++++++++++++++++++++
\begin{comment}
\end{comment}
%++++++++++++++++++++++++++++++++++++++++++++++++++++++++++++++++++++++

Two decades ago, empirical evidence concerning the existence and frequency of 
planets around stars, other than our own, was absent. Since this time, the 
detection of extrasolar planets from Jupiter-sized to most recently Earth-sized 
worlds has blossomed and we are finally able to shed light on the plurality of 
Earth-like, habitable planets in the cosmos. Extrasolar moons may also be 
frequent habitable worlds but their detection or even systematic pursuit remains 
lacking in the current literature. Here, we present a description of the first 
systematic search for extrasolar moons as part of a new observational project 
called ``The Hunt for Exomoons with \emph{Kepler}'' (HEK).
The HEK project distills the entire list of known transiting planet candidates
found by \emph{Kepler} (2326 at the time of writing) down to the most
promising candidates for hosting a moon. Selected targets are fitted using
a multimodal nested sampling algorithm coupled with a planet-with-moon
light curve modelling routine. By comparing the Bayesian evidence of a 
planet-only model to that of a planet-with-moon, the detection process is 
handled in a Bayesian framework. In the case of null detections, upper limits 
derived from posteriors marginalised over the entire prior volume will be 
provided to inform the frequency of large moons around viable planetary hosts, 
$\eta_{\leftmoon}$. After discussing our methodologies for target selection,
modelling, fitting and vetting, we provide two example analyses.

\end{abstract}

% #####################################################################
%% keywords

%% EOF keywords
%% EOF titlepage

% #####################################################################

%%%%%%%%%%%%%%%%%%%%%%%%%%%%%%%%%%%%%%%%%%%%%%%%%%%%%%%%%%%%%%%%%%%%%%%%%%%%%%%%
%%%%%%%%%%%%%%%%%%%%%%%%%%%%%%%%%%%%%%%%%%%%%%%%%%%%%%%%%%%%%%%%%%%%%%%%%%%%%%%%
%%%%%%%%%%%%%%%%%%%%%%%%%%%%%%%%%%%%%%%%%%%%%%%%%%%%%%%%%%%%%%%%%%%%%%%%%%%%%%%%

\section{INTRODUCTION}
\label{sec:intro}

Extrasolar moons represent an outstanding challenge in modern observational
astronomy. Their detection and study would yield a revolution in the 
understanding of planet/moon formation and evolution, but perhaps most 
provocatively they could be frequent seats for life in the Galaxy 
\citep{williams:1997}. Their existence is widely speculated both from a 
Copernican perspective and simulations of planet synthesis \citep{elser:2011},
and yet their detection has remained elusive. 

In truth, very little effort has been spent on the search for moons relative to 
their planetary counterparts, presumably due to the known difficulty such a feat 
represents. Indeed, there has never been a systematic search for exomoons and 
just a handful of papers have attempted to detect their presence in several
transiting planet systems \citep{brown:2001,kippingbakos:2011a,kippingbakos:2011b} 
(\citealt{lewis:2011} provide a extensive discussion regarding non-transiting 
moons around pulsar planets). With \emph{Kepler} now boasting 2326 candidate 
transiting planets \citep{batalha:2011}, we can consider this candidate list to 
be comparable to target list of stars used in radial velocity surveys, for 
example. Further, \emph{Kepler} is specifically designed to detect Earth-sized 
transiting objects and thus has the capability to find large moons 
\citep{kipping:2009}. In light of this, the time is ripe for the first 
systematic program to search for the satellites of extrasolar planets and so in 
this paper we describe our new observational project: ``The Hunt for Exomoons 
with \emph{Kepler}'' (HEK).

The objective of the HEK project is to inspect the most viable planetary 
candidates for evidence of an exomoon, using publicly available \emph{Kepler}
photometry. In cases of a null-detection, and when conditions permit, the 
obtained constraints will be used to inform a new statistic in exoplanetary 
science: the frequency of large moons, $\eta_{\leftmoon}$.

The structure of the paper is as follows.
% 2. EXOMOONS
In \S\ref{sec:exomoons}, we will present an overview of the current literature 
regarding the existence of exomoons, which will inform our search. 
% 3. OBS EFFCTS
In a similar vein, \S\ref{sec:obseffects} presents an overview of the current
literature on the observational consequences of exomoons.
% 4. TARGET SELECTION
In \S\ref{sec:targets}, the multi-pronged HEK target selection process will be 
described, which greatly aids in refining the search to only the most favourable 
objects.
% 5. FITTING
\S\ref{sec:fitting} discusses the detection, modelling and fitting strategy of
the HEK project, with particular focus on the use of Bayesian evidence and
the application of a multimodal nested sampling algorithm, \multi, in
combination with our exomoon-modelling code, \luna.
% 6. VETTING
In \S\ref{sec:vetting}, we explore the process of vetting exomoon candidates
against false-positives, such as perturbing planets.
% 7. WORKED EXAMPLES
\S\ref{sec:examples} presents two examples of the \multi\ algorithm with \luna\
on synthetic data, illustrating the use of Bayesian evidence for detections.
% 8. SUMMARY
Finally, we summarise the project's goals and methods in \S\ref{sec:summary}.
A list of acronyms used throughout this work is available in the Appendix,
Table~\ref{tab:acros}. Similarly, a list of mathematical symbols and notation
used in this work is found in Table~\ref{tab:parameters}.

%%%%%%%%%%%%%%%%%%%%%%%%%%%%%%%%%%%%%%%%%%%%%%%%%%%%%%%%%%%%%%%%%%%%%%%%%%%%%%%%
%%%%%%%%%%%%%%%%%%%%%%%%%%%%%%%%%%%%%%%%%%%%%%%%%%%%%%%%%%%%%%%%%%%%%%%%%%%%%%%%
%%%%%%%%%%%%%%%%%%%%%%%%%%%%%%%%%%%%%%%%%%%%%%%%%%%%%%%%%%%%%%%%%%%%%%%%%%%%%%%%

\section{EXOMOONS}
\label{sec:exomoons}

\subsection{Definition}
\label{sub:definition}

An extrasolar moon, or exomoon, is a natural satellite of a extrasolar planet.
Our project will focus on the detection of satellites which are gravitationally
bound to be within the Hill sphere of their host planet, as opposed to
quasi-satellites which can reside outside. %Therefore the focus of our project
%is clearly on moons as opposed to Trojans.

For a planet with a single large moon, it is possible that a more appropriate 
description would be a binary-planet. The blurry line between a binary-planet
and a true planet-moon pair can be drawn by the point at which the centre of
mass of two bodies lies outside the radius of both bodies. This distinction
is merely a taxonomical issue though and in principle the HEK project can also 
detect binary planets, should they exist.

\subsection{Observational Effects of an Exomoon}
\label{sub:obs}

A detailed discussion of how we will search for exomoons is presented in
\S\ref{sec:obseffects}. However, for the purposes of providing some context in 
the subsequent sections, we will briefly outline the observational consequences 
of an exomoon. The HEK project will search for moons around transiting 
extrasolar planets and thus primarily make use of photometry (specifically
photometry obtained by \emph{Kepler}). For an overview of alternative exomoon 
detection techniques, see \citet{thesis:2011}. For transiting planet systems, a
companion moon can reveal itself through two categories of observational 
effects:

\begin{itemize}
\item[{\tiny$\blacksquare$}] Dynamical variations of the host planet.
\item[{\tiny$\blacksquare$}] Eclipse features induced by the moon.
\end{itemize}

Dynamical effects are measured as perturbations of the motion of the 
host planet away from a simple Keplerian orbit. It is thought that the most
observable dynamical effects will be transit timing and duration variations
(TTV and TDV respectively) \citep{sartoretti:1999,szabo:2006,kipping:2009a,
kipping:2009b}. Dynamical effects primarily reveal information about the 
exomoon mass. 

Eclipse features are caused by the moon either occulting the stellar light
directly or inducing so-called ``mutual events'' \citep{ragozzine:2010} by
occulting the planet during a planet-star eclipse. In either case, such events
primarily reveal information about the exomoon radius.

\subsection{Feasibly Detectable Systems}
\label{sub:detectable}

\citet{kipping:2009} conducted a feasibility study of \emph{Kepler}'s 
sensitivity to a habitable-zone gas giant with a single moon, based upon 
dynamical effects only (i.e. TTV \& TDV). The study assumed moons were on 
coplanar, circular orbits around their host planet. They found that 
\emph{Kepler} should be sensitive to exomoons of masses 
$\gtrsim0.2$\,$M_{\oplus}$, in the most favourable circumstances. Even in this 
optimistic case, this is an order-of-magnitude more massive than the most 
massive moon in our Solar System, specifically Ganymede at 
$0.025$\,$M_{\oplus}$. Consequently, it must be understood that HEK is searching 
for moons in a mass regime which do not exist within our own Solar System. We
dub these moons as ``large moons'' and place a definition of such a moon as
being $\gtrsim 10^{-1}$\,$M_{\oplus}$. Again, this is an arbitrary distinction 
but is useful for the HEK project.

We point out that HEK will be more sensitive than the limits in 
\citet{kipping:2009} due to the fact that we will also search for eclipses 
caused by the moon. This additional information means that we must be at least 
as sensitive as using dynamics alone, but should become significantly more 
sensitive by including this extra information. An in-depth analysis of our 
sensitivity to exomoons is beyond the scope of this work. We consider such an 
analysis unnecessary in light of the fact that each system studied will have 
upper limits derived and so the detectability of moons will become apparent as 
the project develops. At this time, we consider a far more useful application of 
our efforts to be in actually conducting the search.

\subsection{Plausible Origin of Large Moons}
\label{sub:origin}

Given that no large moons ($M_S\gtrsim10^{-1}$\,$M_{\oplus}$) exist in the Solar 
System, how likely is it that such an object exists? The first important point 
to make is that two classes of moons exist: regular satellites and irregular 
satellites. In discussing the plausibility of a large moon around a planet, 
there are two issues to consider: i) origin ii) evolution. In other words, the 
moon must have some initial plausible formation/origin scenario and second it 
must survive long enough to be detected. We here discuss the former issue.

Regular satellites are those which are believed to have formed in-situ around
the host planet as it accumulates gas and rock-ice solids from solar orbit.
\citet{canup:2006} predict that the cumulative moon mass is constrained via 
$\sum M_{S,i} \lesssim 2\times10^{-4}$\,$M_P$, where $M_P$ is the mass of the 
planet host. This limit is caused by a balance of two competing processes; the 
supply of inflowing material to the satellites, and satellite loss through 
orbital decay driven by the gas. Thus, for Jupiter-like masses, large moons 
should not form. Although brown-dwarfs could harbour Earth-mass moons, the mass 
ratio remains too small for TTV/TDV methods to feasibly infer their presence. 
Therefore, we argue it is unlikely HEK will detect any regular satellites if the 
\citet{canup:2006} scaling law holds.

Irregular satellites are those which are obtained from a non-local origin such
as capture (e.g. Triton; \citealt{agnor:2006}) or impact scenarios (e.g. the 
Moon; \citealt{taylor:1992}). Such moons could reach large masses so long as 
they are dynamically stable. This means that Earth-mass moons (or even larger) 
are plausible and HEK would be sensitive to such objects. Unlike regular 
satellites, irregular moons may frequently reside in retrograde orbits too 
\citep{porter:2011}.

In order for a planet to capture a large moon, an essentially terrestrial-mass 
object must be dynamically captured by a larger body. The probability of such
an event occurring is a topic of active theoretical research and HEK will
elevate it to a topic of active observational research too. Several
plausible capture mechanisms have been proposed. From a case study of Triton,
\citet{agnor:2006} propose that this moon was originally a member of a binary
which encountered Neptune during its migration through the proto-Kuiper belt.
Upon encountering Neptune, a momentum-exchange reaction occurred ejecting one
member and bounding the other as a satellite. One difficulty with this mechanism
is that a binary terrestrial planet pair is required to endow an Earth-mass moon
around a gas giant and the abundance of such binaries is unclear.

Another possible capture process is via atmospheric tunneling, where a
terrestrial mass object encounters the atmosphere of the gas giant, inducing
strong drag effects leading to large changes in momentum \citep{williams:2011}. 
The obvious extreme limit of this scenario is an impact between two rocky cores, 
where the smaller mass body is broken up and later reforms in a close-in orbit, 
as is proposed for the Moon's origin \citep{taylor:1992}. Impacts do seem to be 
a feasible way of forming moons with \citet{elser:2011} recently simulating the 
interaction of planetesimals and finding that Earth-Moon pairs should be 
common.

For planets which do not migrate through a proto-Kuiper belt or under the
assumption that such objects will never reach sufficient mass to qualify as
large moons, an alternative source of terrestrial mass objects is required. This
object could be an inner terrestrial planet encountered during the gas giant's
inward migration or even a large, unstable Trojan which librates too close
to the planet. Indeed, \citet{eberle:2011} have shown that a gas giant planet 
(in their case HD 23079b) can capture an Earth-mass Trojan into a stable 
satellite orbit, occurring in 1 out of the 37 simulations they ran.

Any capture process (as opposed to an impact) will tend to produce very 
loosely-bound initial orbits, leading us to question how many of these captured
bodies may ultimately survive. \citet{porter:2011} have investigated this 
issue and found that captured moons have encouraging survival rates, with a 
survival probability of order 50\% for various planet-moon-star configurations.

Finally, true binary planets are in principle also plausible. 
\citet{podsiadlowski:2010} showed that one viable scattering history in the 
formation of a planetary system is the tidal capture of two planets forming a 
binary. Indeed, a Jupiter-Earth pair could be considered as an extreme binary, 
much like Pluto-Charon.

\subsection{Plausible Evolution of Large Moons}
\label{sub:evolution}

With several plausible avenues for a large moon to become bound to a planet,
the next requirement is that the moon can survive long enough to be observed.
\citet{porter:2011} show that captured moons tend to start in eccentric,
inclined orbits and rapidly circularise and relax into a coplanar orbit. With
a moon on a circular, coplanar orbit within the Hill sphere, the moon evolves
by either spinning-in or out through tidal effects\footnote{Tides can also
induce significant heating on the satellite \citep{cassidy:2009}}. If the 
rotational period of the planet is shorter than the moon's orbital period, the 
tides cause the moon to spin-out over time. The opposite process occurs if this 
ratio is reversed \citep{barnes:2002}.

Regardless as to whether the moon moves inwards or outwards, the spatial
boundary conditions must be given by the moon being in contact with the planet
(minimum spatial separation) to being outside the sphere of gravitational 
influence of the planet (the maximum stable separation). The time to move
between these two limits is insensitive to the direction of the moon's 
migration \citep{barnes:2002}. A fortuitous moon could in principle 
reverse the direction of its migration just before hitting one of these spatial
limits (due to braking of the planet's rotation for example) and thus 
essentially double its maximum allowed lifetime. However, we do not consider 
such a scenario to be likely. Since the tidal dissipation depends upon the mass 
of the moon, one can write the maximum allowed exomoon mass as a function of the 
moon's lifetime (expression taken from \citealt{barnes:2002}). This yields

\begin{align}
M_{S,\mathrm{max}} &= \frac{2}{13} \Bigg( \frac{\mathfrak{D}_{\mathrm{max}}^3 a_{B*}^3}{3 M_*} \Bigg)^{13/6} \frac{M_P^{8/3} Q_P}{3 k_{2p} \mathbb{T} R_P^5 \sqrt{G} },
\label{eqn:maxmoon}
\end{align}

where $M_*$ is the stellar mass, $M_P$ is the planetary mass, $Q_P$ is the tidal 
quality factor of the planet, $k_{2p}$ is the Love number of the planet, $G$ is 
the gravitational constant, $R_P$ is the planetary radius, $a_{B*}$ is the 
orbital semi-major axis of the planet-moon barycentre around the host star and 
$\mathbb{T}$ is the lifetime of the moon. Equation~\ref{eqn:maxmoon} suggests 
that Earth-like (i.e. habitable-zone) moons are plausible around Jupiters for
 billions of years around stars of mass $M_* > 0.4$\,$M_{\odot}$ 
\citep{barnes:2002}.

In Equation~\ref{eqn:maxmoon}, $\mathfrak{D}_{\mathrm{max}}$ represents the 
maximum stable semi-major axis for the moon, in units of the Hill radius. One 
may naively expect this should be equal to unity but in reality three-body 
perturbations tend to disrupt a moon before it reaches the Hill radius. Through 
detailed numerical integrations, \citet{domingos:2006} have shown that 
$\mathfrak{D}_{\mathrm{max}} = 0.4895$ for prograde satellites and 
$\mathfrak{D}_{\mathrm{max}} = 0.9309$ for retrograde satellites (assuming 
circular orbits).

Finally, a planet which migrates inwards will eventually lose its
moon(s) due to the shrinking Hill sphere.  The Hill radius is
given by

\begin{align}
R_H &= a_{B*} \Bigg(\frac{M_P}{3 M_*}\Bigg)^{1/3}.
\label{eqn:hillradius}
\end{align}

One can therefore see that the Hill radius decreases linearly with the orbital
semi-major axis of the host planet. \citet{namouni:2010} showed that since
planetary migration occurs much faster than moon migration, a moon
initially well-inside the Hill radius can quickly find itself outside. 
\citet{namouni:2010} estimate moons tend to be lost by this process for 
$a_{B*} \lesssim 0.1$\,AU.

The planetary parameters clearly have a significant impact on the survivability
of a large moon. Since in general the planetary parameters may be reasonably
estimated based upon the transit light curve, it is possible to create a list
of the most favourable planetary candidates for exomoon inspection. More details 
on this target selection procedure are given in \S\ref{sec:targets}.

\subsection{Objectives of HEK}
\label{sub:objectives}

The existence of large moons is hypothetically plausible, but currently we have 
no empirical evidence to test this hypothesis. For this reason, the objectives
of the HEK project will be as follows:

\begin{enumerate}
\item The primary objective of HEK is to search for signatures of extrasolar
moons in transiting systems.
\item The secondary objective of HEK will be to derive posterior distributions,
marginalised over the entire prior volume, for a putative exomoon's mass and 
radius, which may be used to place upper limits on such terms (where conditions 
permit such a deduction).
\item The tertiary objective of HEK is to determine $\eta_{\leftmoon}$- the 
frequency of large moons bound to the Kepler planetary candidates which could 
feasibly host such an object (in an analogous manner to $\eta_{\oplus}$ - the 
frequency of Earth-like planets).
\end{enumerate}

For the primary objective, the issue of upper limits or detection biases is
irrelevant and in many ways this is the simplest task to execute on candidate
systems. In contrast, the secondary objective requires more care due to
the inter-parameter correlations. For example, \citet{kippingbakos:2011b} show
how the excluded 3-$\sigma$ limit on a putative moon around TrES-2b is strongly 
correlated to the assumed semi-major axis and inclination of the moon. For this
reason, any upper limits must relate a posterior marginalised over the entire
prior volume (more details on this are given in \S\ref{sub:fitting}). 
Additionally, in some cases it may not be possible to provide a mass or radius 
upper limit if, for example, a strong eclipse signal is found but ultimately 
deemed to be a false positive. Finally, the tertiary objective is challenging in 
light of the numerous detection biases which plague any survey such as this. For
example, two problematic biases come from the fact we preferentially select 
systems with short-cadence data (see \S\ref{sub:TSP}) and visual anomalies (see 
\S\ref{sub:TSV}) in the light curve.

In the case of a definitive signal, we require a detection significance
threshold. Currently, 2326 planetary candidates are known \citep{batalha:2011},
but this number is constantly increasing. We may conservatively set 
our false-alarm-probability threshold to be 1 in 10,000 or 3.89-$\sigma$. This 
is rounded up to 4-$\sigma$ for our nominal detection threshold. Statistically,
this implies that 1 in every 15,787 claimed detections will be false, although
in reality the false-positive-rate is a function of how carefully we vet our
candidate signals rather than merely the number of sigmas the signal is detected
to. Acknowledging this, we will follow the set of detection criteria defined in 
\citet{thesis:2011}. Most relevant of these, the exomoon targets must be 
physically feasible solutions (criterion C3). Details on our vetting process, 
which will undoubtedly evolve as the HEK project develops, are given in 
\S\ref{sec:vetting}.

In the case of a null-detection, the objective of HEK will be determine the 
exomoon mass and radius which can be excluded, which will aid in the 
determination of $\eta_{\leftmoon}$. As already mentioned, these upper limits
will be based on a posterior marginalised over the entire prior volume. Using
a 4-$\sigma$ upper limit is usually not very informative since they allow for
``hidden moons'', for example a moon which is behind the planet in every 
transit epoch (e.g. see \S\ref{sub:nulleg}). In such a case, the radius limit is 
essentially unbound. We therefore choose to give the lower constraint of 90\% 
confidence upper limits. Although less stringent, these limits are more useful 
in a population-sample of exomoon candidates. Further, all null-detections will 
have the posterior distributions made publicly available on the project website 
(www.cfa.harvard.edu/HEK/) so that the community may investigate 
$\eta_{\leftmoon}$ using our results.

\subsection{Why HEK is Possible}
\label{sub:whyHEKpossible}

HEK is feasible for two reasons. Firstly, as already discussed, \emph{Kepler}
has yielded extraordinary success with 2326 transiting candidates down to
Earth-sized objects \citep{batalha:2011}. Secondly, recent advancements in the 
theoretical development of exomoon-search methods make a search feasible with 
modern computational resources. Specifically, the \luna\ algorithm 
\citep{luna:2011} offers a completely analytic and exact solution for the 
planet-moon light curve including dynamics, non-linear limb darkening and the 
modelling of mutual events. Methods based upon even partial numerical 
implementation (such as pixelating the star) dramatically impinge our ability to 
run light curve fits, given the large number of parameters and prior volume
which must be explored in any given fit. Thus, \emph{Kepler} and \luna\ are both 
critical to the feasibility of HEK.

%%%%%%%%%%%%%%%%%%%%%%%%%%%%%%%%%%%%%%%%%%%%%%%%%%%%%%%%%%%%%%%%%%%%%%%%%%%%%%%%
%%%%%%%%%%%%%%%%%%%%%%%%%%%%%%%%%%%%%%%%%%%%%%%%%%%%%%%%%%%%%%%%%%%%%%%%%%%%%%%%
%%%%%%%%%%%%%%%%%%%%%%%%%%%%%%%%%%%%%%%%%%%%%%%%%%%%%%%%%%%%%%%%%%%%%%%%%%%%%%%%

\section{OBSERVATIONAL EFFECTS}
\label{sec:obseffects}

\subsection{Dynamical Variations}
\label{sub:dynamicalvariations}

As discussed in \S\ref{sub:obs}, there exists two broad categories of 
observational consequences of an exomoon in the transit light curve: dynamical
variations and eclipse features. In this section, we will overview these 
techniques, each of which is a key tool to the HEK project. We begin our
discussion with dynamical variations, of which there are several flavours.

\subsubsection{Transit timing variations (TTV)}

The first conceived effect was transit timing variations (TTV), by 
\citet{sartoretti:1999}, which is conceptually analogous to the astrometric 
technique of finding planets around stars. The motion of the planet around the 
planet-moon barycentre causes a planet to transit periodically early and late. 
For co-aligned, circular orbits, \citet{sartoretti:1999} showed that the maximum 
deviation in the time between two transits would be

\begin{align}
\Delta t &\sim \frac{ \mathfrak{D} M_S P_{B*} }{ 3^{1/3} \pi M_P^{2/3} M_*^{1/3} } \nonumber \\
\qquad &= 36.0\,\mathfrak{D}\,\Big(\frac{M_S}{M_{\oplus}}\Big)\,\Big(\frac{P_{B*}}{\mathrm{years}}\Big) \Big(\frac{M_J}{M_P}\Big)^{2/3} \Big(\frac{M_{\odot}}{M_*}\Big)^{1/3}\,\mathrm{mins}.
\label{eqn:sar99scaled}
\end{align}

where $P_{B*}$ is the orbital period of the planet-moon baycentre around the 
star (the ``B'' subscript will be used to refer to the planet-moon barycentre
throughout). Encouragingly, planets on orbital periods of $\sim~1$ year could 
generate very significant TTV amplitudes. However, \citet{kipping:2009a} pointed 
out two critical hurdles with the technique. Firstly, many other effects can 
cause TTVs aside from moons, notably perturbing planets 
\citep{agol:2005,holman:2005}, and there seemed to be no obvious way to 
discriminate between the physical source of the measured TTV. Secondly, the 
satellite's period around the barycentre ($P_{SB}$) is less than the planet's 
orbital period for all bound orbits. Specifically, \citet{kipping:2009a} have 
shown that

\begin{align}
P_{SB} &= P_{B*} \sqrt{\frac{\mathfrak{D}^3}{3}},
\label{eqn:kiprelation}
\end{align}

since $\mathfrak{D}<1$ for all bound exomoons, the period of the exomoon will 
always be less than 60\% of the planet's period. This means the TTV waveform 
is undersampled\footnote{This applies to not only TTV, but for all 
exomoon-induced timing effects i.e. TDV-V and TDV-TIP} since one can only 
measure a timing deviation once per transit. This undersampling means that one 
cannot reliably infer the period of the exomoon signal (only a set of harmonic 
solutions) and instead one can only reliably  measure the RMS amplitude (i.e.
the scatter) which scales as $\sim a_{SB} M_S$. Whilst knowing $a_{SB} M_S$ is 
useful, clearly knowledge of each component would be more powerful in 
understanding an exomoon.

\citet{thesis:2011} generalized the original \citet{sartoretti:1999} equation 
for the amplitude of the TTV effect, to account for longitude of the ascending 
node, inclinations, eccentricities and position of pericentres (see 
\citealt{thesis:2011} for definitions of the various terms). The 
root-mean-square (RMS) amplitude of the full TTV effect is given by

\begin{align}
\delta_{\mathrm{TTV}} &= \frac{a_{SB} M_S P_{B*}}{a_{B*} M_P} \frac{(1-e_{SB}^2) \sqrt{1-e_{B*}^2}}{(1+e_{B*}\sin\omega_{B*})} \sqrt{\frac{\Phi_{\mathrm{TTV}}}{2\pi}}, %\\
\label{eqn:TTVrms}
\end{align}

where the $\Phi$ term contains information on the eccentricity
and can be found in \citet{thesis:2011}. The associated waveform is given by

\begin{align}
\mathrm{TTV} &\simeq \Bigg[\frac{a_{SB} \sqrt{1-e_{B*}^2} (1-e_{SB}^2) M_S P_{B*}}{2 \pi a_{B*} M_P (1+e_{B*}\sin\omega_{B*})}\Bigg] \Lambda_{\mathrm{TTV}},
\end{align}

where

\begin{align}
\Lambda_{\mathrm{TTV}} &= [1+e_{SB}\cos \nu_{SB}]^{-1} [\cos\varpi_{SB} \cos(\nu_{SB}+\omega_{SB}) \nonumber \\
\qquad& -\sin i_{SB} \sin \varpi_{SB} \sin(\nu_{SB}+\omega_{SB})].
\label{eqn:TTVwaveform}
\end{align}

where $\nu_{SB}$ is the true anomaly of the satellite around the planet-moon
barycentre.

\subsubsection{Velocity induced transit duration variations (TDV-V)}

TDV-V was conceived of a decade after TTV by \citet{kipping:2009a}.
\citet{kipping:2009a} showed that the same motion responsible for changes
in position causing TTV should also cause changes in velocity. Since the 
planet's velocity is inversely proportional to the duration of the transit, 
TDV-V must be another observational consequence of extrasolar moons. Whereas TTV 
scales as $a_{SB} M_S$, thus favouring the detection of moons at large 
separation, TDV-V scales as $a_{SB}^{-1/2} M_S$ and so favours the detection of 
close-in moons. TDV-V is conceptually analogous to the radial velocity method of 
finding planets (except really it is \emph{tangential} velocity).

Besides from the complementarity of their parameter space coverage, TTV and 
TDV-V could also yield a unique solution for $M_S$ and $a_{SB}$, if both effects 
were detected \footnote{$P_{SB}$ may then may easily calculated through Kepler's 
Third Law}. This therefore solves the undersampling issue presented by only
detecting one of the two effects, as discussed earlier for TTV. Further 
more, for coplanar, circular moons, the two effects exhibit a phase difference 
of $\pi/2$ radians since the dynamics is essentially projected simple harmonic 
motion. This phase difference offers a method to distinguish an exomoon TTV 
from, say, a perturbing planet induced TTV. Therefore, detecting both TTV and 
TDV-V solves the issue of solution uniqueness (assuming the moon is coplanar and 
circular) and mass determination. Defining the $\tilde{T}_B$ as the duration
for the planet-moon barycentre to enter then exit the stellar disc, the TDV-V 
waveform may be described via \citep{thesis:2011} to be

\begin{align}
\mathrm{TDV-V} &= \tilde{T}_B \Bigg(\frac{a_{SB} M_S P_{B*}}{a_{B*} M_P P_{SB}}\Bigg) \nonumber \\
\qquad& \times \Bigg(\frac{\sqrt{1-e_{B*}^2}}{\sqrt{1-e_{SB}^2} (1+e_{B*}\sin\omega_{B*})}\Bigg) \Lambda_{\mathrm{TDV-V}},
\end{align}

where

\begin{align}
\Lambda_{\mathrm{TDV-V}} &= \cos\varpi_{SB} [e_{SB}\sin\omega_{SB}+\sin(\nu_{SB}+\omega_{SB})] \nonumber \\
\qquad& + \sin i_{SB} \sin\varpi_{SB} [e_{SB}\cos\omega_{SB}+\cos(\nu_{SB}+\omega_{SB})].
\label{eqn:TDVwaveform}
\end{align}

The correponding RMS amplitude may be found through integration over $\nu_{SB}$
(see \citealt{thesis:2011} for details) and yields

\begin{align}
\delta_{\mathrm{TDV-V}} &= \tilde{T}_B \Bigg(\frac{a_{SB} M_S P_{B*}}{a_{B*} M_P P_{SB}}\Bigg) \nonumber \\
\qquad& \times \Bigg(\frac{\sqrt{1-e_{B*}^2}}{\sqrt{1-e_{SB}^2} (1+e_{B*}\sin\omega_{B*})}\Bigg) \sqrt{\frac{\Phi_{\mathrm{TDV-V}}}{2\pi}}, %\\
\label{eqn:TDVrms}
\end{align}

where the $\Phi$ term again contains information on the eccentricity
and can be found in \citet{thesis:2011}.

\subsection{Transit impact parameter induced transit duration variations 
(TDV-TIP)}

%%% TDV-TIP EXPLAN
\begin{figure}
\begin{center}
\includegraphics[width=8.4 cm]{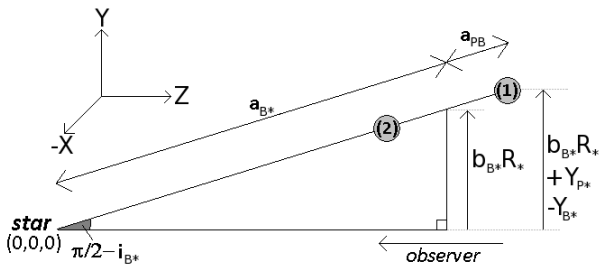}
\caption{\emph{Cartoon of the TDV-TIP effect. Here, the moon is relaxed 
into the same orbital plane as the planet's orbit and causes the planet to 
experience reflex motion illustrated by the two positions of the planet, (1) and 
(2). This motion can be seen to cause a change in the apparent impact parameter, 
which causes a change in the transit duration.}} 
\label{fig:tipexplan}
\end{center}
\end{figure} 

Later, \citet{kipping:2009b} showed that a second transit duration variation 
(TDV) effect exists, dubbed TDV-TIP, standing for transit-impact-parameter 
induced TDV. The physical origin of this effect is that if the planet-moon 
orbital plane is not precisely normal to the sky (which is practically always 
true), then the planet's reflex motion must yield a component orthogonal to both 
the observer's line-of-sight and the tangent to the planetary motion. This 
motion leads to periodic variations in the apparent transit impact parameter. 
Since the transit duration is a strong function of the impact parameter 
\citep{seager:2003}, then duration variations must follow. The TDV-TIP 
effect is typically much smaller than the TDV-V (velocity) effect but is 
strongly enhanced for near-grazing transits. Intrigueingly, TDV-TIP allows one 
to determine the sense of orbital motion of the moon (i.e. prograde or 
retrograde) and thus could be a key tool in understanding the origin of the 
satellite (although such cases require high signal-to-noise). The phase of 
TDV-TIP is constructive to TDV-V for prograde orbits and destructive for 
retrograde. The TDV-TIP waveform is shown in \citet{thesis:2011} to be

\begin{align}
\mathrm{TDV-TIP} &= \tilde{T}_B \Bigg( \frac{b_{B*}}{1-b_{B*}^2} \Bigg) \Bigg( \frac{a_{SB} M_S (1-e_{SB}^2)}{R_* M_P}\Bigg) \Lambda_{\mathrm{TDV-TIP}},
\end{align}

where

\begin{align}
\Lambda_{\mathrm{TDV-TIP}} &= (1+e_{SB}\cos \nu_{SB})^{-1} \Bigg[ \sin(\nu_{SB}+\omega_{SB}) \nonumber \\
\qquad& \times [-\cos i_{SB} \sin i_{B*} + \sin i_{SB} \cos i_{B*} \cos\varpi_{SB}] \nonumber \\
\qquad& + \cos i_{B*} \sin\varpi_{SB}\cos(\omega_{SB}+\nu_{SB}) \Bigg].
\label{eqn:TIPwaveform}
\end{align}

And the corresponding RMS amplitude is given by

\begin{align}
\delta_{\mathrm{TDV-TIP}} &= \tilde{T}_B \Bigg( \frac{b_{B*}}{1-b_{B*}^2} \Bigg) \nonumber \\
\qquad& \times \Bigg( \frac{a_{SB} M_S (1-e_{SB}^2)}{R_* M_P}\Bigg) \sqrt{\frac{\Phi_{\mathrm{TDV-TIP}}}{2\pi}}, %\nonumber \\
\label{eqn:TIPrms}
\end{align}

where the $\Phi$ term again contains information on the eccentricity
and can be found in \citet{thesis:2011}.

It should noted that due to the undersampling issue discussed in regard to
TTV, TDV-V and TDV-TIP are simply observed as a global TDV effect. 
Thus we have two timing observables, TTV and TDV.

\subsection{Eclipses Features}
\label{sub:eclipsefeatures}

\begin{figure*}
\begin{center}
\includegraphics[width=16.8 cm]{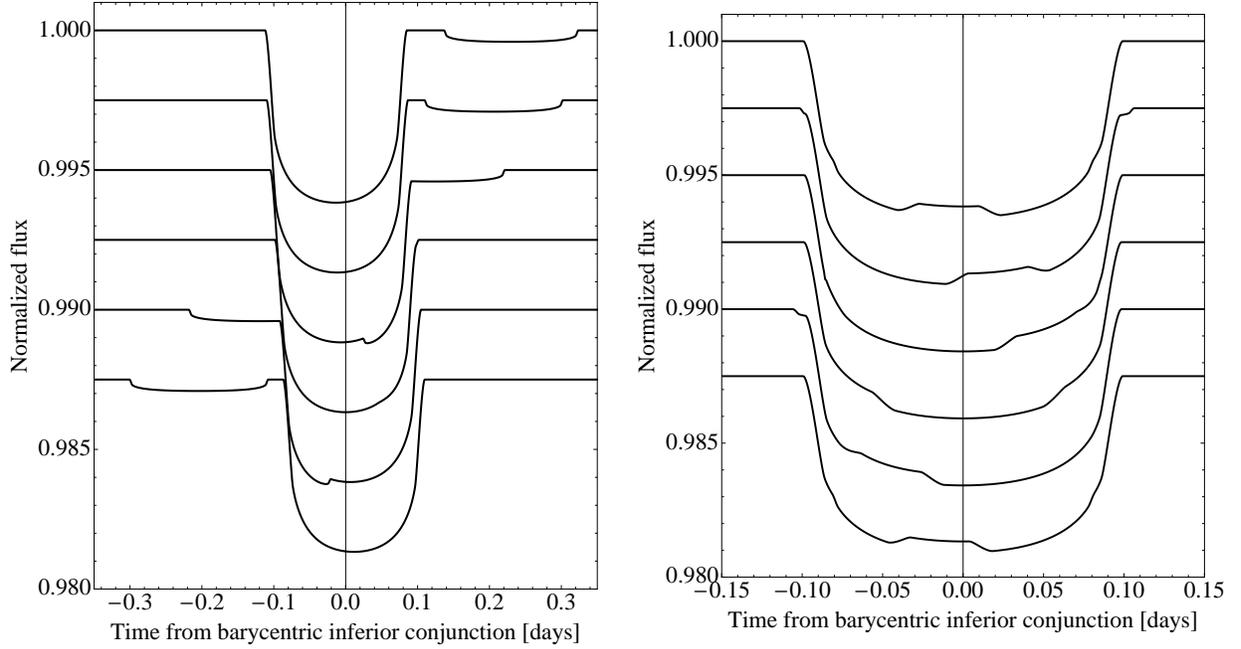}
\caption{\emph{Left panel: transit light curves of a planet with a moon on a 
wide separation, demonstrating auxiliary transits. Right panel: transit light 
curves of a planet with a close-in moon, demonstrating mutual events. See 
\citet{luna:2011} for details of the parameters used in these simulations 
(Figures 7\&6 respectively).}} 
\label{fig:eclipses}
\end{center}
\end{figure*}

A second class of observational effect we can search for is the eclipse of the 
moon. This eclipse can be either in-front of the star or in-front/behind the 
planet during the planet-star transit. The former, which we refer to as
``auxiliary transits'', is more likely to be detected for moons on wide orbits. 
The latter, sometimes dubbed ``mutual events'' \citep{ragozzine:2010,pal:2011}, 
is geometrically more probable for moons on close-in orbits. We show examples of 
each of these types in Figure~\ref{fig:eclipses}.

%%% ECLIPSE EXPLAN

Critically, unlike TTV and TDV effects, eclipses are sensitive to the exomoon
radius (and not the mass). This presents the opportunity of determining
the density of the moon. Further, in an analogous manner to how transiting 
planets may yield the mean stellar density, $\rho_*$, \citep{seager:2003}, 
transiting moons allow one to measure the mean planetary density directly, 
$\rho_P$, as first shown in \citet{weighing:2010}. Armed with both $\rho_*$ and 
$\rho_P$ plus the ratio of the $R_P/R_*$, one can determine $M_P/M_*$ 
\citep{weighing:2010}. This powerful trick means that even if no TTV or TDV 
signal is detected, the eclipses of the moon alone allow one to measure the mass 
of the exoplanet (assuming $M_*$ is known). If radial velocities exist for the 
system, $M_*$ can be measured directly through this technique 
\citep{weighing:2010}. Additionally, the dynamically determined $M_P/M_*$ can be 
compared to the RV solution using a-priori value of $M_*$ (say from 
spectroscopy) to ensure consistency.

The ability to weigh planets using moons plays a key role in the HEK project. 
Candidates with unphysical properties can be quickly rejected as plausible exomoon 
signals.

%%%%%%%%%%%%%%%%%%%%%%%%%%%%%%%%%%%%%%%%%%%%%%%%%%%%%%%%%%%%%%%%%%%%%%%%%%%%%%%%
%%%%%%%%%%%%%%%%%%%%%%%%%%%%%%%%%%%%%%%%%%%%%%%%%%%%%%%%%%%%%%%%%%%%%%%%%%%%%%%%
%%%%%%%%%%%%%%%%%%%%%%%%%%%%%%%%%%%%%%%%%%%%%%%%%%%%%%%%%%%%%%%%%%%%%%%%%%%%%%%%

\section{TARGET SELECTION}
\label{sec:targets}

\subsection{Candidate vs Planet}
\label{sub:plancandidates}

At the timing of writing, 2326 \emph{Kepler} transiting candidate 
planets have been announced \citep{batalha:2011}. However, the public
candidate list is only 1235 at the time of writing \citep{borucki:2011}. HEK
will employ the most up-to-date list during the development of the project.
It should also be noted that included within the candidate list are several 
candidates which have already been confirmed as bona-fide planets to a high 
degree of confidence e.g. the Kepler-9 system \citep{torres:2011}.

Despite the progress made in confirming these objects through novel
techniques, such as BLENDER \citep{torres:2004}, the vast majority of
the planetary candidates remain unconfirmed.

One possible approach might therefore be to only inspect the confirmed
planets for exomoons. However, such a strategy is unnecessarily 
restrictive. The detection of an exomoon would in fact allow one
to measure the mass and radius of the planet, as well
as the moon \citep{weighing:2010}. By modelling these systems, one
essentially obtains the same information which is normally provided
by radial velocity (RV) or multi-planet transit timing variations (TTV) 
- the masses\footnote{Stictly speaking, one obtains the mass ratios}. 
Therefore, exomoon detection is a planetary confirmation tool, in addition to 
the RV and TTV techniques. Consequently, there is no need to limit ourselves to 
confirmed planets only and the HEK project will consider all planetary 
candidates as possible exomoon hosts.

\subsection{Target Selection (TS) Overview}
\label{sub:focus}

Equipped with the capacity of exomoons to confirm planetary candidates, 
analyzing all of the hundreds of planetary candidates for evidence of exomoons 
would be the most comprehensive way to proceed. However, practical limitations, 
such as man-power and computational constraints, make such a task unrealistic. 
Therefore, we must conduct a target selection (TS) phase. Target selection works 
by taking the full list of planetary candidates (in our case the KOIs presented
in \citealt{borucki:2011}) and identifying those targets which are of 
highest priority for more intensive investigation. We have three strategies for
target selection, which have some overlap:

\begin{enumerate}
\item Visual inspection a subset of planetary candidates for exomoon-like 
eclipse features (TSV). 
\item Automatic filtering of the planetary candidates, based upon system
parameters (TSA). 
\item Targets of opportunity (TSO).
\end{enumerate}

In some cases, a planetary candidate may fall into more than one category. The
identified candidates are then further prioritized by hand. This final stage,
which we call target selection prioritization (TSP), is typically done by 
experience of what constitutes a feasible candidate (see \S\ref{sub:TSP} for
more information).

\subsection{Visual Target Selection (TSV) Overview}
\label{sub:TSV}

Visual Target Selection (TSV) is typically conducted by first selecting
sub-sample of planetary candidates. This subset is typically selected from
some simple parameter constraints such as planet size and semi-major axis. The 
aim is to allow this subset to have only a weak theoretical prejudice for where 
to expect a moon. The subset sample size may range from dozens to hundreds of 
light curves. The light curves are then inspected for features resembling 
exomoon-like signals, notably:

\begin{itemize}
\item[{\tiny$\blacksquare$}] Mutual events, preferably with a 
flat-top (to discriminate from a starspot crossing) e.g. right-panel of 
Figure~\ref{fig:eclipses}.
\item[{\tiny$\blacksquare$}] Auxiliary transits (a second distinct transit 
feature offset from the primary) e.g. left-panel of Figure~\ref{fig:eclipses}.
\item[{\tiny$\blacksquare$}] Repeating anomalies in the light curve of any
morphology.
\end{itemize}

The visual inspection of \emph{Kepler} photometry has been already been shown
to be a useful technique for finding transiting planet candidates 
\citep{fischer:2012,lintott:2012} and here we extend such inspections to hunting 
for moons too. Whereas TSA relies heavily on the derived parameters of the 
system (plus an assumed mass-radius relation), TSV is far more ignorant of these 
values. The benefit of this is that we are able to catch interesting systems 
which would go missed by TSA due to potential errors in the assumed system
parameters (for example from the Kepler Input Catalogue, KIC) or even our 
theoretical understanding of where moons may reside. TSV systems are later 
scrutinized to see whether the signal is sufficiently high signal-to-noise, in 
particular relative to any time correlated noise in the photometry, but this 
forms part of our final prioritization stage, TSP (see \S\ref{sub:TSP} for more
information).

\subsection{Automatic Target Selection (TSA) Overview}
\label{sub:focus}

TSA works by applying a set of filters which reduce the total planetary
candidate list to to a more manageable size. There are several considerations 
affecting the target selection:

\begin{itemize}
\item[{\tiny$\blacksquare$}] Availability - Systems for which there exists 
available photometry.
\item[{\tiny$\blacksquare$}] Reliability - Systems for which we have reasonable 
parameters.
\item[{\tiny$\blacksquare$}] Capability - Systems which are capable of hosting 
large moons.
\item[{\tiny$\blacksquare$}] Detectability - Systems which are capable of 
presenting a detectable moon signal.
\end{itemize}

TSA leans heavily on the system parameters. Notably, the planetary mass must be
essentially guessed based upon the radius. Details on this procedure are 
presented in \S4.4.2.

Stellar parameters of the planetary candidates have been estimated by the Kepler 
team as part of the Kepler Input Catalogue (KIC), a photometric survey of stars 
in the \emph{Kepler} field-of-view designed to identify bright dwarfs as targets 
for the mission. \citet{brown:2011} describes the estimation of physical 
parameters, whereby the synthetic spectra of \citet{castelli:2004} are 
forward-modelled with effective temperature, $T_{\mathrm{eff}}$, surface 
gravity, log(g), and metallicity, log(Z), as free parameters to match the 
photometric measurements in the KIC. A relation between luminosity, effective 
temperature and surface gravity, derived from the stellar evolutionary models 
computed by \citet{girardi:2000}, is used to estimate the stellar masses, which, 
when combined with log(g), estimates the radii. 

\citet{brown:2011} state that the KIC effective temperature and radius 
estimations are reliable for Sun-like stars, but are ``untrustworthy'' for stars 
with $T_{\mathrm{eff}}<3750$\,K. To accommodate this possible weakness, TSA
is run on both the KIC-only catalogue and a catalogue of KIC plus that of 
\citet{muirhead:2011}, who use near-infrared spectroscopy. These 
cases will be flagged appropriately. By repeating for both catalogues, we make
no decision about which catalogue is the correct one and ensure we do not
miss any interesting objects. Any other catalogues which appear will be
treated in a similar manner.

\subsubsection{Availability}

At the time of writing, the only publicly available photometry from 
\emph{Kepler} comes from quarters 0, 1, 2 \& 3 (Q0, Q1, Q2 \& Q3). In all cases, 
Q1 (33.5d), Q2 (88.7d) and Q3 (89.3d) photometry exists and naturally this
will continue to expand as time goes on. Only in some cases is 
Q0 (9.3d) photometry available. This is because Q0 was technically
``calibration'' data and originally not expected to be useful 
scientifically. As a result, the number of sources observed was only 
52,496 in Q0 but expanded to 156,097 for Q1. Whilst some sources were 
dropped from the target list, the majority were kept and thus roughly a 
third of all sources have Q0 photometry available, in addition to the other
quarters.

In most instances, photometry data is currently available in long-cadence (LC) 
mode only, whereas SC is preferable for our search for reasons described in 
\S\ref{sub:TSP}. There are three possible exceptions to this rule:

\begin{enumerate}
\item A planetary candidate was detected very early on and thus the data was
able to be switched to short-cadence (SC) for subsequent observations.
\item The star already had a known exoplanet and thus was observed in
SC from the outset.
\item The star was coincidentally pre-selected for asteroseismology and 
thus was observed in SC for one or more of the three available quarters.
\end{enumerate} 

Case 3) is not very useful to us because the subset is rather small. Case
2) is also not useful because the known exoplanets, HAT-P-7b, HAT-11-b
and TrES-2b, are all short-period hot-Jupiters which are unlikely to be
good exomoon hosts. Case 1) does not guarantee SC data because only 512
targets can be observed in this mode at any one time and the number of
planetary candidates exceeds this value significantly (2326; 
\citealt{batalha:2011}).

In conclusion, it is generally unlikely that a potentially interesting target 
(for exomoon hunting) will have any useful SC data available at the start of the
HEK project and this represents a major limitation on our capabilities. Those 
that do are highly valuable to HEK, during this stage of the project, given 
that SC data strongly enhances our capability to detect exomoons. As we 
mentioned before though, more quarters are becoming public over time, including 
potentially much more SC data, and thus the effectiveness of our program is 
likely to improve significantly over time.

To look for exomoons, we require systems where at least three transits have been 
observed. This is because three transits represents the minimum number 
for timing deviations to be inferred. Two transits only introduces a 
total degeneracy between the orbital period of the planet and the 
transit timing variation (TTV) amplitude. Let us denote the total baseline of 
data as $B$\,days and assume optimistically that Q0 exists. In order to 
``guarantee''\footnote{This is not strictly a guarantee due to the potential
for data gaps but nevertheless serves as a useful minimum requirement.} three 
transits have been recorded, we set the maximum allowed orbital period to be 
$P_P \leq B/4$, which is our first filter.

\subsubsection{Reliability}

\emph{Kepler} measures the ratio-of-radii of planetary candidates, rather
than their absolute radii. From photometric measurements discussed in
\citet{brown:2011}, the \emph{Kepler} input catalogue (KIC) contains estimated 
masses and radii for all of the host stars. Therefore, the planetary radii can 
be determined to a reasonable degree of reliability.

However, one of the key parameters in both predicting the feasibility of 
a moon around a planet and the detectability is the mass of the host 
planet, $M_P$ (particularly for computing the Hill sphere). \emph{Kepler} 
cannot measure the masses of these objects, except in some extreme cases where 
ellipsoidal variations \citep{mislis:2012} or TTVs exist \citep{agol:2005}.

The maximum allowed exomoon mass around a planet scales as 
$M_{S,\mathrm{max}} \sim M_P^{8/3}$ (see Equation~\ref{eqn:maxmoon}). If we 
overestimate this value, we will waste time looking at systems where moons 
cannot exist. If we underestimate this value, we will miss some potentially 
interesting candidates. However, the number of candidates is large and 
we can afford to miss some objects for the sake of focussing on the very best
candidates. Therefore, we choose to provide a minimum estimate of $M_P$.

Planets are frequently broken up into three regimes: 
i) Super-Earths ii) Neptunes iii) Jupiters. For a Super-Earth, the 
mass-scaling relationship is arguably the most predictable relative to the other 
two regimes, due to the known properties of rock under pressure. The precise
location of the boundary between a gas/ice giant and a rocky Super-Earth, and 
the conditions under which such a boundary varies, is empirically a subject in 
its infancy. However, models of the growth of a protoplanet's envelope suggest
a critical mass of $M_P \sim 20$\,$M_{\oplus}$ (see review article 
\citealt{angelo:2010}), whereafter the envelope growth timescale rapidly 
diminishes by orders of magnitude. This suggests that the maximum radius of a 
rocky planet is $R_P \sim 2$\,$R_{\oplus}$, using the scaling relation of
\citet{valencia:2006}. This boundary is consistent with that adopted in other
works, such as \citet{borucki:2011}. Therefore, if a planet has 
$R_P\lesssim2\,R_{\oplus}$, we consider it more likely to be icy/rocky rather 
than gaseous. The assumed mass of the Super-Earth is calculated using the
\citet{valencia:2006} relation $(R_P/R_{\oplus}) \sim (M_P/M_{\oplus})^{0.27}$.
We stress that this assumption is only used for target selection and is not 
adopted in the actual fits or system analyses.

\begin{figure}
\begin{center}
\includegraphics[width=8.4 cm]{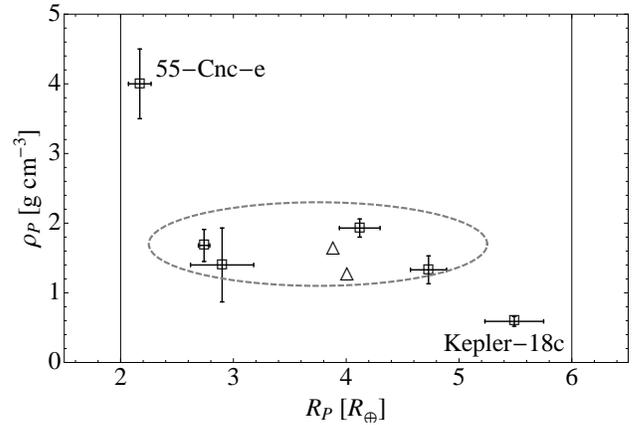}
\caption{\emph{Bulk densities of all known transiting exoplanets in the 
``Neptune'' regime. We only show planets with densities determined with
SNR$>3$. Triangles mark the position of Neptune and Uranus for context.
Vertical grid lines mark the Super-Earth and Jupiter boundaries, as defined by
\citet{borucki:2011}. We mark the location of the somewhat anomalous 55-Cnc-e
and Kepler-18c. The gray ellipse represents a 2-$\sigma$ bound of the bulk of 
points yielding an approximately uniform density of $1.7\pm0.3$\,g\,cm$^{-3}$.}} 
\label{fig:Nep_densities}
\end{center}
\end{figure}

The definition of a Neptune is somewhat more tenuous but we use the same
definition as that of \citet{borucki:2011} - planets which satisfy
$2\,R_{\oplus}<R_P<6\,R_{\oplus}$. Transiting ``Neptunes'' are in short supply 
in the exoplanet literature, especially those with well-determined densities
(SNR$>3$)\footnote{We define SNR as the reported value for the density divided
by its uncertainty}. Figure~\ref{fig:Nep_densities} shows bulk density versus 
the radius of the six exoplanets which satisfy this criteria. Systems parameters 
are taken from \citet{gillon:2011}, \citet{kundurthy:2011}, \citet{fossey:2012},
\citet{bakos:2009}, \citet{henry:2011} and \citet{cochran:2011} for
55-Cnc-e, GJ 1214b, GJ 436b, HAT-P-11b, HD 97658b and Kepler-18c respectively.

From these, four out of the six settle around an approximately common
density close to that of Neptune and Uranus of 
$\bar{\rho}_P = 1.7 \pm 0.3$\,g\,cm$^{-3}$ (where the uncertainty is the 
standard deviation of these four). The median of all six is 1.5\,g\,cm$^{-3}$. 
55-Cnc-e has a much higher density of $4.0_{-0.3}^{+0.5}$\,g\,cm$^{-3}$, 
possibly due to its close proximity (in radius) to the Super-Earth/Neptune 
boundary ($R_P = 2.17 \pm 0.10$\,$R_{\oplus}$) and is ostensibly not a typical 
member of the $2\rightarrow6$\,$R_{\oplus}$ category. Similarly, Kepler-18c is 
the largest radii planet ($R_P = 5.5\pm0.3$,$R_{\oplus}$) close to the
Neptune/Jupiter boundary of 6\,$R_{\oplus}$ and exhibits a much lower density of 
$0.6\pm0.1$\,g\,cm$^{-3}$, and thus may not be typical either.

We therefore decide to settle on an assumed density of
$1.7$\,g\,cm$^{-3}$ to get an assumed planetary mass in TSA. The function of 
this assumed mass will be primarily in computing the maximum allowed exomoon 
mass and a generous margin will be assigned to this calculation to account for 
the fact we have made such a broad approximation (as will be discussed later). 
As before, this again only used for TSA and not employed in the final analyses.

%%% Neptune densities

Planets above the $6\,R_{\oplus}$ boundary show much greater variation in 
density, going from extrema of $0.122_{-0.042}^{+0.072}$\,g\,cm$^{-3}$ for 
WASP-17b \citep{anderson:2010} to $26.4\pm5.6$\,g\,cm$^{-3}$ for CoRoT-3b
\citep{deleuil:2008}. At this point, we consider the reliability of any 
mass-radius relation to be untenable and thus we add the filter than any planet 
of radius $R_P>6$\,$R_{\oplus}$ are not considered. This also helps in
the exomoon detection too, since smaller planets present larger mutual 
events with a putative companion \citep{luna:2011}.

It is important to stress that in all cases that the assumed masses are only
used for target selection purposes and will not be adopted as final
values in the actual system analyses.

\subsubsection{Capability}

For each \emph{Kepler} candidate, we need to know the capability of the planet
to host a moon. The maximum allowed exomoon mass from Equation~\ref{eqn:maxmoon}
\citep{barnes:2002} offers a useful way of accomplishing this. We assume
a retrograde moon so that the maximum allowed semi-major axis of the moon's 
orbit is 93.09\% of the Hill radius (i.e. $\mathfrak{D}_{\mathrm{max}}=0.9309$).

To compute the maximum allowed moon mass, we use the system parameters 
presented in \citet{borucki:2011} or more recent results where available. As
discussed earlier, the planetary mass is assumed following
some simple scaling rules. Finally, we assume the same Jovian-like tidal 
dissipation parameters as that used by \citet{barnes:2002}:
specifically $Q_P = 10^5$ and $k_{2p} = 0.51$. The $k_{2p}$ value is that for 
a $n=1$ polytrope \citep{hubbard:1984} and the $Q_P$ factor is consistent with
estimates for Jupiter \citep{goldreich:1966}. This then allows us to compute
the maximum allowed exomoon mass, assuming $\mathbb{T}=5$\,Gyr, using 
Equation~\ref{eqn:maxmoon}, taken from \citet{barnes:2002}.

It is important to realize this calculation is intended to only point 
the way towards potentially interesting targets. It is not intended to 
be the final determination of this value, which is simply not possible 
with the current information available.

Accordingly, we exclude all candidates for which the maximum moon mass is 
below $10$\,$M_{\oplus}$. This high limit is deliberately an overestimate to
allow for the number of assumptions and approximations made thus far. We further 
exclude planets with $a_{B*}<0.1$\,AU based upon the arguments given in 
\S\ref{sub:evolution} and \citep{namouni:2010}.

\subsubsection{Detectability}

In order to make an exomoon detection, we require a mass and radius. The timing
amplitudes, which yield the exomoon mass, depend upon the configuration of the
system but the eclipse amplitude, yielding the exomoon radius, has a far weaker 
correlation to these terms. The eclipse amplitude is approximately given by
$(R_S/R_*)^2$ and thus for any given target we should be able to easily
estimate the eclipse signal of an Earth-sized moon i.e. $(R_{\oplus}/R_*)^2$.

We may also estimate the SNR over a 6.5-hour integration time by simply 
dividing the exomoon eclipse depth by the CDPP values presented in 
\citet{borucki:2011}. We filter all out results where SNR$<1$ for a 
single event.

\subsection{Opportunity Target Selection (TSO) Overview}

Additionally, we will consider ``targets of opportunity'' for special objects of
interest. We envision these will typically be confirmed, published \emph{Kepler} 
exoplanets. These targets offer numerous advantages in that the entire
photometric time series used in the discovery paper is usually available (i.e. 
many more quarters than normal), SC data is often available, the planet is 
known to be genuine and frequently follow-up information such as spectroscopy, 
radial velocities or even asteroseismology are usually available too. We 
envision that these TSO targets would be typically selected for HEK analysis 
because they have special significance (e.g. are in the habitable-zone).

\subsection{Target Selection Prioritization (TSP) Overview}
\label{sub:TSP}

The prioritization stage is the process of selecting just a few targets out of
the candidates found by the TSA, TSV and TSO stages. TSP is typically done
using detailed light curve inspection, experience of what constitutes a viable
signal and other factors.

For example the availability of short-cadence (SC) data is a key TSP factor,
due to the improved sensitivity relative to long-cadence (LC) data. This is 
because many of the exomoon features induced on the light curve can occur on 
timescales shorter than 30\,minutes and thus would be lost in the LC data. 
Further, SC data yields higher resolution of the ingress/egress of the planetary 
transit which thus yields a tighter determination of the planetary parameters 
\citep{binning:2010}. With lower uncertainty on the planetary signal, we are 
naturally able to more easily distinguish exomoon signals. Therefore, targets 
with even partial SC data are strongly preferred to those with exclusively LC 
photometry.

%%%%%%%%%%%%%%%%%%%%%%%%%%%%%%%%%%%%%%%%%%%%%%%%%%%%%%%%%%%%%%%%%%%%%%%%%%%%%%%%
%%%%%%%%%%%%%%%%%%%%%%%%%%%%%%%%%%%%%%%%%%%%%%%%%%%%%%%%%%%%%%%%%%%%%%%%%%%%%%%%
%%%%%%%%%%%%%%%%%%%%%%%%%%%%%%%%%%%%%%%%%%%%%%%%%%%%%%%%%%%%%%%%%%%%%%%%%%%%%%%%

\section{FITTING}
\label{sec:fitting}

\subsection{Modelling Strategy: \luna}
\label{sub:luna}

Modelling the eclipses of planet-moon systems is non-trivial, if one wishes
to retain analytic expressions. The advantages of an analytic model are
manifold, allowing CPU intensive fitting techniques (e.g. Monte Carlo methods) 
to fully explore the complex parameter space. The requirement for an analytic
algorithm excludes the methods presented in \citet{simon:2009}, 
\citet{sato:2009}, \citet{deeg:2009}, and \citet{tusnski:2011}.

The most significant challenge, in terms of analytic modelling, is when the 
planet, moon and star all partially overlap. The analytic solution for the area 
of overlap of three circles was only recently found by \citet{fewell:2006} and 
this discovery allowed for the first time an exomoon code which could be totally 
analytic in nature.

\citet{luna:2011} presented an algorithm to this end, dubbed \luna, which
dynamically models the planet-moon motion and utilizes the \citet{fewell:2006}
solution (plus numerous new solutions derived in \citealt{luna:2011}) to
produce simulated light curves for moons. The code uses quadratic limb
darkening and runs almost as fast as generating a planet signal by itself
(i.e. the code of \citealt{mandel:2002}). As a result, \luna\ is easily
implemented with Monte Carlo based fitting techniques. Further, the dynamical 
component of \luna\ means that effects such as TTV, TDV-TIP and TDV-V are all 
inherently accounted for, plus other previously unconsidered effects such as 
ingress/egress asymmetry. \luna\ is a potent weapon in exomoon detection.

We note that another analytic algorithm capable of modelling exomoon eclipses
has appeared recently in \citet{pal:2011}. However, the HEK project will
make use of \luna\ alone, since this already satisfies all of our requirements.

\luna\ also features several other light curve analysis techniques developed
recently, such as accounting for the finite integration time using selective 
resampling \citep{binning:2010} and accounting for blended/third-light using
the methodology of \citet{kiptin:2010}.

\subsection{Detection Strategy: Bayesian Model Selection}
\label{sub:bayesian}

The process of making a detection of any physical phenomenon is essentially an
exercise in model selection. In our case, this is most simply described by
comparing how well the data are explained by a planet-only model (the null
hypothesis) versus a planet-with-moon model.

The Bayesian framework is a very powerful basis for these model comparisons,
allowing the observer to incorporate prior knowledge (such as the allowed
physical bounds of various parameters) and naturally include ``Occam's razor'' 
as a way of penalizing overly-complicated models. Whilst various information
criterion have been proposed for performing model selection, the use
of Bayesian evidence has emerged as the metric of choice to perform model
comparisons \citep{liddle:2007}.

Bayesian inference methods provide a consistent approach to the estimation of a 
set parameters $\mathbf{\Theta}$ in a model $\mathcal{M}$ for the data 
$\mathbf{D}$. Bayes' theorem states that

\begin{align}
\mathrm{Pr}(\mathbf{\Theta}|\mathbf{D},\mathcal{M}) &= \frac{\mathrm{Pr}(\mathbf{D}|\mathbf{\Theta},\mathcal{M}) \mathrm{Pr}(\mathbf{\Theta}|\mathcal{M}) }{\mathrm{Pr}(\mathbf{D}|\mathcal{M})},
\end{align}

where $\mathrm{Pr}(\mathbf{\Theta}|\mathbf{D},\mathcal{M}) = \mathcal{P}(\mathbf{\Theta})$ 
is the	posterior probability distribution of the parameters, 
$\mathrm{Pr}(\mathbf{D}|\mathbf{\Theta},\mathcal{M}) = \mathcal{L}(\mathbf{\Theta})$ 
is the likelihood, $\mathrm{Pr}(\mathbf{\Theta}|\mathcal{M}) = \pi(\mathbf{\Theta})$ 
is the prior, and $\mathrm{Pr}(\mathbf{D}|\mathcal{M}) = \mathcal{Z}$ is the 
Bayesian evidence.

The evidence can be understood to be simply the factor required to normalize the 
posterior over $\mathbf{\Theta}$, so that

\begin{align}
\mathcal{Z} &= \int \mathcal{L}(\mathbf{\Theta})\pi(\mathbf{\Theta})\,\mathrm{d}^D\mathbf{\Theta},
\label{eqn:evidenceintegral}
\end{align}

where $D$ is the dimensionality of the parameter space. Since the evidence is
the average of the likelihood over the prior, Occam's razor is inherently
included. Therefore, a simpler theory with more compact parameter space will
yield a larger evidence than a more intricate theory. Model selection between
two competing theories, $\mathcal{M}_0$ and $\mathcal{M}_1$ can be decided by
comparing their respective posterior probabilities, for the given data, via

\begin{align}
\frac{\mathrm{Pr}(\mathcal{M}_1|\mathbf{D})}{\mathrm{Pr}(\mathcal{M}_0|\mathbf{D})} &= \frac{\mathcal{Z}_1}{\mathcal{Z}_0} \frac{\mathrm{Pr}(\mathcal{M}_1)}{\mathrm{Pr}(\mathcal{M}_0)},
\end{align}

where $\mathrm{Pr}(\mathcal{M}_1)/\mathrm{Pr}(\mathcal{M}_0)$ is the a-priori
probability ratio for the two models, typically set to unity but occasionally
requires more thought. In this way, the odds ratio of a planet-with-moon model 
can be assessed, relative to a planet-only model. Defining the Bayes' factor as 
$\mathcal{B} = \mathcal{Z}_1/\mathcal{Z}_0$, $|\log \mathcal{B}| > 6$ indicates
a $>3$\,$\sigma$ detection, $|\log \mathcal{B}| > 10$ indicates $>4$\,$\sigma$
and $|\log \mathcal{B}| > 15$ indicates $>5$\,$\sigma$. 

\subsection{Fitting Strategy: \multi}
\label{sub:fitting}

In the exoplanet literature, Markov Chain Monte Carlo (MCMC) techniques have
emerged as the favoured tool for fitting transit light curves. However, the
technique is primarily for parameter estimation and does not natively yield
the Bayesian evidence. Whilst techniques such as thermodynamic integration
(e.g. see \citealt{ruan:1996}) can get round this problem, it comes at the
cost of great computational expense.

\subsubsection{Nested sampling}

Nested sampling \citep{skilling:2004} is a Monte Carlo method which puts the 
calculation of the Bayesian evidence in a central role, but also produces 
posterior inferences as a by-product. Nested sampling is generally considerably
more efficient than MCMC methods. For example, in cosmological applications, 
\citet{mukherjee:2006} showed that their implementation of the method requires a 
factor of $\sim100$ fewer posterior evaluations than thermodynamic integration.

A full discussion of nested sampling is given in \citet{skilling:2004} and
\citet{feroz:2008}. We here provide a brief description, following the
notation of \citet{feroz:2008}, for the purposes of conceptually illustrating 
the technique.

\begin{figure}
\begin{center}
\subfigure[]{\includegraphics[width=0.4\columnwidth]{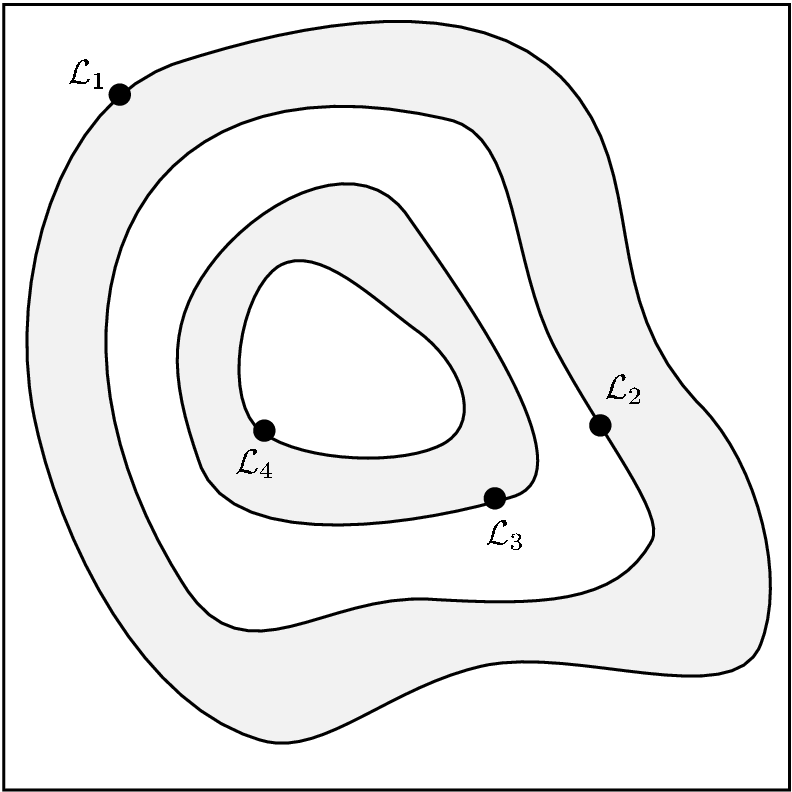}}\hspace{0.3cm}
\subfigure[]{\includegraphics[width=0.4\columnwidth]{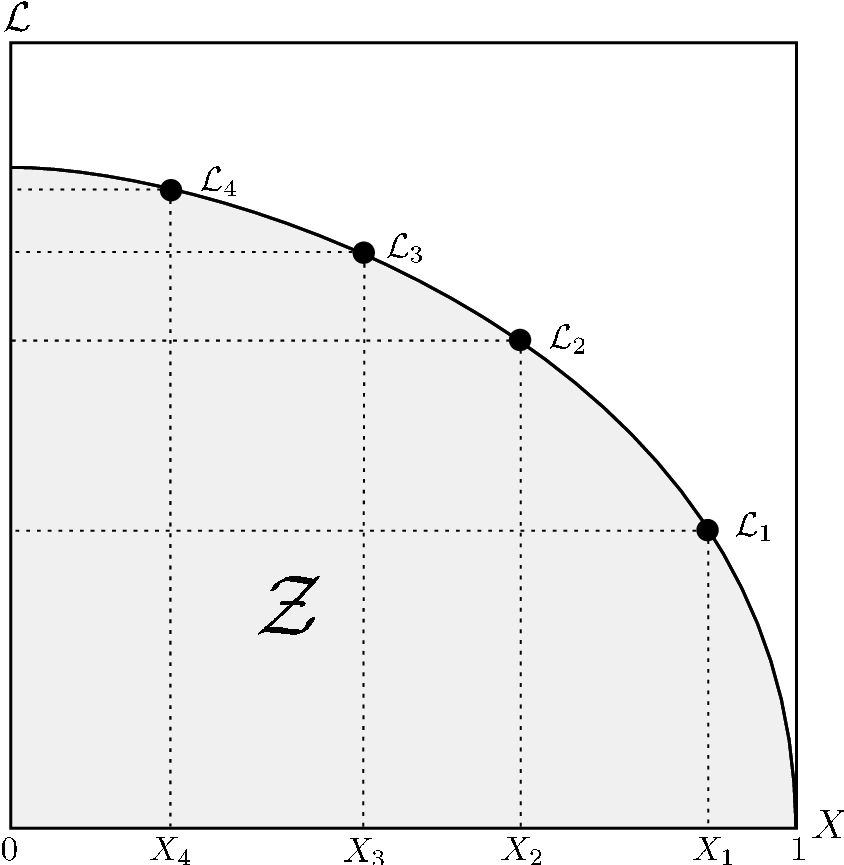}}
\caption{Cartoon illustrating (a) the posterior of a two dimensional problem; 
and (b) the transformed $\mathcal{L}(X)$ function where the prior volumes 
$X_{i}$ are associated with each likelihood $\mathcal{L}_{i}$.}
\label{fig:nestedcartoon}
\end{center}
\end{figure}

Nested sampling takes advantage of the relationship between the likelihood and 
the prior volume to transform the multidimensional evidence integral 
(Equation~\ref{eqn:evidenceintegral}) into a more manageable one-dimensional 
integral. The ``prior volume'' X is defined by 
$\mathrm{d}X = \pi(\Theta)\mathrm{d}^D\Theta$, such that

\begin{align}
X(\lambda) = \int_{\mathcal{L}(\mathbf{\Theta})>\lambda} \pi(\mathbf{\Theta})\,\,\mathrm{d}^D\mathbf{\Theta},
\label{eqn:priorvolume}
\end{align}

where the integral extends over the region(s) of parameter space contained
within the iso-likelihood contour $\mathcal{L}(\mathbf{\Theta}) = \lambda$. One 
may then re-write the evidence integral of Equation~\ref{eqn:evidenceintegral} 
in the form

\begin{align}
\mathcal{Z} = \int_{0}^{1} \mathcal{L}(X)\,\,\mathrm{d}X,
\end{align}

where $\mathcal{L}(X)$, which is the inverse of Equation~\ref{eqn:priorvolume},
is a monotonically decreasing and continuous function of $X$. Consequently,
if one can evaluate a series of $M$ likelihoods via 
$\mathcal{L}_i = \mathcal{L}(X_i)$, where $X_i$ is a sequence of decreasing 
volumes from unity ($X_0$) to zero ($X_M$) as shown schematically in 
Figure~\ref{fig:nestedcartoon}, then the evidence may be approximated 
numerically as a weighted sum using

\begin{align}
\mathcal{Z} = \sum_{i=1}^M \mathcal{L}_i w_i.
\end{align}

The weights may be computed using the trapezium rule, 
$w_i = \frac{1}{2} (X_{i-1} - X_{i+1})$. The algorithm works by casting a net
of $N$ ``active'' points across the initial prior space. The active point with 
the lowest likelihood is removed (made ``inactive'') and a new replacement point 
is generated such that its likelihood is higher than this rejected value. As the 
algorithm progresses, it travels through nested shells of likelihood as the 
prior volume is reduced. The routine is terminated once the evidence is computed 
to some tolerance precision, typically 0.5 in log-evidence. Upon termination,
parameter posteriors may also be computed using the active and inactive points.

\subsubsection{Multimodal nested sampling}

Multimodal nested sampling is an implementation of nested sampling to both
account for multiple modes and achieve efficient sampling via the use of the
``simultaneous ellipsoidal sampling'' method, described in \citet{feroz:2007}.
A publicly available version of the algorithm, dubbed \multi, is
available. We direct the reader to \citet{feroz:2008} for a description of
the algorithm and its use (including several toy examples).

To date, applications of the technique have been mostly limited to cosmology,
gravitational wave detection and particle physics (e.g. see 
\citealt{vegetti:2010}, \citealt{feroz:2009} and \citealt{abdussalam:2010}
respectively). Recently however, \citet{feroz:2011} demonstrated the first 
application of the technique to radial velocity data for detecting extrasolar 
planets. Here, we briefly discuss our application of \multi\ to transit 
light curve fitting (we note that a detailed study of \multi\ for 
fitting transits is currently in preparation by Balan et al., personal 
communication).

\subsubsection{Combining \multi\ and \luna}

Nested sampling allows one to compute the Bayesian evidence of a model fit
at a much lower computational cost than using thermodynamic integration
with MCMC techniques. However, there are many other advantages of employing
\multi\ with \luna, rather than an MCMC.

The most simple and common flavour of MCMC is the Metropolis-Hastings algorithm,
widely used in the recent exoplanet literature. With a unimodal likelihood
function, this technique is effective and robust, capable of identifying the
single minimum even when the initial starting point of the chain is widely
separated from this minimum. However, multimodal distributions are extremely
problematic. If the spacing between two modes is much greater than the width of
the proposal distribution, then the MCMC will take an inordinate amount of time
to cross over and practically speaking the MCMC is stuck in a local minimum.
For a planet-only model, a unimodal likelihood distribution is generally 
expected but a planet-with-moon model exhibits many modes, especially due to
harmonic power in the TTVs and TDVs \citep{kipping:2009a}. 

\multi\ comprehensively searches the entire prior volume identifying all minima
and thus is not affected by this problem. We point out, however, that more 
elaborate flavours of MCMC can still find the global minimum but at further
computational cost. Techniques such as parallel tempering \citep{geyer:1991},
differential evolution \citep{braak:2006} and genetic crossovers 
\citep{gregory:2009} can be appended into the MCMC methodology. 

To implement \luna\ with \multi, we simply need to define a likelihood function.
\multi\ calls \luna\ for each active point and \luna\ returns a likelihood
value of the point. It is only at this point where communication between the
two algorithms occurs and so is the only outstanding problem in implementing
a combination of \multi\ with \luna. For quiet stars, the noise properties
of the photometry is nearly perfectly Gaussian \citep{kippingbakos:2011b}. For
Gaussian uncorrelated noise, the likelihood function may be simply defined as

\begin{align}
\mathcal{L}(\mathbf{\Theta}) &= \prod_{i=1}^N \frac{1}{\sqrt{2\pi\sigma_i^2}} \exp\Big[-\frac{(f_{\mathrm{obs},i} - f_{\mathrm{mod},i}(\mathbf{\Theta}))^2}{2\sigma_i^2}\Big],
\end{align}

where $f_{\mathrm{obs},i}$ is the observed flux, 
$f_{\mathrm{mod},i}(\mathbf{\Theta})$ is the model flux returned by \luna, and 
$\sigma_i$ is the photometric uncertainty.

In cases where correlated noise constitutes a significant component of the noise
budget, then the likelihood function may be modified accordingly. For example,
\citet{carter:2008} present a wavelet-based likelihood function to compensate
for time-correlated noise.

\subsection{Choosing the Parameters}
\label{sub:parameters}

The parameter set for the fitting procedure should be physically-motivated so
that well-defined boundary conditions may be imposed, have a uniform prior and
exhibit as small as possible mutual correlations with the other fitting terms.
Unless the algorithm is imposed with non-uniform priors, the choice of 
parameters often defines the choice of priors too. 

\subsubsection{Planet parameters}

The choice of the parameter set for fitting a planet-only model is a subject 
which has been investigated in numerous papers in the exoplanet literature
(e.g. see \citealt{carter:2008} and \citealt{investigations:2010}). The
transit of a spherical planet on a circular orbit across a uniformly bright,
spherical star is described by just four parameters, which is expected given the 
near-trapezoidal nature of the light curve. These parameters are 
$\{\tau_{B*},p,(a_{B*}/R_*),b_{B*}\}$, where $\tau_{B*}$ is the time of the 
transit minimum of the planet's barycentre across the star, $p$ is the 
ratio-of-radii (i.e. $R_P/R_*$), $(a_{B*}/R_*)$ is the orbital semi-major axis 
of the planet's barycentre around the star in units of the stellar radius and 
$b_{B*} = (a_{B*}/R_*)\cos i_{B*}$ (where $i_{B*}$ is the orbital inclination of
the planet's barycentre around the star). Multiple transits allows one to fit 
for the orbital period, $P_{B*}$, as well. Of these terms, only $(a_{B*}/R_*)$ 
and $b_{B*}$ are problematic, due to their very strong mutual correlation. 
Further, whilst $b_{B*}$ is bound to be $0<b_{B*}<1$ for fully transiting 
planets, $(a_{B*}/R_*)$ lies within the range $0<(a_{B*}/R_*)<\infty$ and is 
therefore not appropriate for a uniformly distributed prior.

\citet{carter:2008} have suggested using transit duration related replacements
for $(a_{B*}/R_*)$ and $b_{B*}$ to reduce the mutual correlation, but these also 
have essentially unbounded upper limits. \citet{map:2012} have suggested 
replacing $(a_{B*}/R_*)$ with $[\rho_{*}^{\mathrm{circ}}]^{2/3}$ (the mean 
stellar density of a spherical star assuming a circular orbit, to the power of 
two-thirds). Besides from reducing the correlation to $b_{B*}$, the bounded 
limits are far better known if we assume the planet is orbiting a main-sequence 
star and impose an upper limit on the eccentricity. Further, the posteriors may 
be used directly to conduct multibody asteroseismology profiling (MAP) analysis, 
which is useful in constraining eccentricity in multiple systems 
\citep{map:2012}. Thus, we choose to use 
$\{\tau_{B*},p,[\rho_{*}^{\mathrm{circ}}]^{2/3},b_{B*},P_{B*}\}$, and adopt

\begin{align}
\rho_*^{\mathrm{circ}} &\simeq \frac{3\pi[(a_{B*}/R_*)^{\mathrm{circ}}]^3}{G P_{B*}^2},
\end{align}

where we have assumed $M_P \ll M_*$ to simplify the expression.
To allow for non-circular orbits, we fit for $\sqrt{e_{B*}} \sin\omega_{B*}$
and $\sqrt{e_{B*}} \cos\omega_{B*}$, which maintain uniform priors in $e_{B*}$ 
and $\omega_{B*}$ but generally exhibit lower mutual correlations than fitting 
for $e$ and $\omega$ directly. Limb darkening may be 
included by adopting a quadratic limb darkening law characterized by $u_1$ and 
$u_2$ limb darkening coefficients. These are constrained to lie within the range
$0<u_1+u_2<1$ and $u_1>0$ \citep{carter:2009}. An upper bound on $u_1$ may be
estimated by inspection of a typical set of coefficients from stellar atmopshere
models. For example, \citet{claret:2000} have a maximum $u_1$ coefficient of
1.4336 across all listed $T_{\mathrm{eff}}$, $\log g$, etc values and across
all bandpasses. We use $u_1<2$ as a conservative upper limit.

Finally, we allow each transit epoch to have its own unique out-of-transit
baseline normalization factor, OOT. These OOT values are represented by the
vector \textbf{OOT} are typically not presented in final results tables, but
are available upon request.

\subsubsection{Moon parameters}

For a planet-with-moon model, a similar set of parameter arises, but with
subtle differences. For example, the inclination of the moon around the
planet-moon barycentre, $i_{SB}$, can be redefined in terms of an impact 
parameter as before (e.g. $b_{SB} = (a_{SB}/R_*) \cos i_{SB}$),
but the $0<b_{SB}<1$ bound no longer applies, rather we have $-\infty<b_{SB}<\infty$.
Inclination follows $0<i_{SB}<\pi$ for prograde moons and $\pi<i_{SB}<2\pi$ for
retrograde moons. Since neither $\arcsin$ nor $\arccos$ are unique over the
range $0$ to $2\pi$ and the light curve is not symmetric at any boundary,
then neither is appropriate for exomoons. Instead, we simply use $i_{SB}$. We
may also use the full range of $0$ to $2\pi$ for the bounds to allow both
prograde and retrograde solutions to be sought.
The same is true for the longitude of the ascending node, $\Omega_{SB}$, which 
is bound by $-\pi/2<\Omega_{SB}<\pi/2$. Allowing for inclined moons is
important since an exoplanet's obliquity is not expected to tidally decay except 
for orbits relatively close to the star \citep{heller:2011}.

Eccentricity terms such as $e_{SB}$ and $\omega_{SB}$ may be fitted either 
directly or using the replacements $\sqrt{e_{SB}}\sin\omega_{SB}$ and 
$\sqrt{e_{SB}}\cos\omega_{SB}$ (which still maintains a uniform prior in both 
terms). The latter option tends to be less correlated and thus preferable.

$P_{SB}$ may be fitted either directly or fitting for $\log P_{SB}$ to impose a
Jeffrey's prior. In either case, the parameter may range from a moon grazing
the planetary surface ($\sim$ hours timescale) to being at exactly one Hill
radius, occurring at $P_{SB} \simeq P_{B*}/\sqrt{3}$ \citep{kipping:2009a}. The 
phase angle of the moon, $\phi_{SB}$, is simply fitted directly within the range 
of $0$ to $2\pi$. $(a_{SB}/R_*)$ raises similar problems as was encountered with 
facing $(a_{B*}/R_*)$. However, we may again make use of the same density trick; 
namely, through Kepler's Third Law we have

\begin{align}
\rho_P &\simeq \frac{3\pi(a_{SB}/R_*)^3}{G P_{SB}^2 p^3},
\end{align}

where we have assumed $M_S \ll M_P$ to simplify the expression.
Physically motivated and sensible bounds on $\rho_P$ may be easily
estimated, in addition to $P_{SB}$ and $p$. It is convenient to fit for the
mass and radius ratios directly using $M_S/M_P$ and $R_S/R_P$, since the
boundary conditions are given by zero to unity in both cases. Since the mass and 
radii ratios are related to the density ratio, we choose to keep the planetary 
density analogous to the stellar density by putting to the power of two-thirds 
again i.e. we fit for $\rho_P^{2/3}$. Table~\ref{tab:priors} summarizes the 
priors used.

\begin{table}
\caption{\emph{Planet-moon parameters used in light curve fits and their 
associated priors.}} % title of Table
\centering % used for centering table
\begin{tabular}{c c} % centered columns (6 columns)
\hline
Parameter & Prior \\ [0.5ex] % inserts table
%heading
\hline
\emph{Planet Parameters} & \\
\hline
$p$ & $\mathcal{U}\{0,0.25\}$ \\
$[\rho_{*}^{\mathrm{circ}}]^{2/3}$\,\,[$\mathrm{kg}^{2/3}$\,$\mathrm{m}^{-2}$] & $\mathcal{U}\{[\rho_{*}^{\mathrm{circ,min}}]^{2/3},[\rho_{*}^{\mathrm{circ,max}}]^{2/3}\}$ \\
$b_{B*}$ & $\mathcal{U}\{0,1\}$ \\
$P_{B*}$\,\,$[\mathrm{days}]$ & $\mathcal{U}\{P_{B*}^{\mathrm{min}},P_{B*}^{\mathrm{max}} \}$ \\
$\tau_{B*}$\,\,[BJD$_{\mathrm{TDB}}$] & $\mathcal{U}\{\tau_{B*}^{\mathrm{min}},\tau_{B*}^{\mathrm{max}}\}$ \\
$\sqrt{e_{B*}}\cos\omega_{B*}$ & $\mathcal{U}\{-1,1\}$ \\
$\sqrt{e_{B*}}\sin\omega_{B*}$ & $\mathcal{U}\{-1,1\}$ \\
$(u_1+u_2)$ & $\mathcal{U}\{0,1\}$ \\
$u_1$ & $\mathcal{U}\{0,2\}$ \\
\textbf{OOT} & $\mathcal{U}\{0.95,1.05\}$ \\
\hline
\emph{Moon Parameters} & \\
\hline
$R_S/R_P$ & $\mathcal{U}\{0,1\}$ \\
$M_S/M_P$ & $\mathcal{U}\{0,1\}$ \\
$[\rho_{P}]^{2/3}$\,\,[$\mathrm{kg}^{2/3}$\,$\mathrm{m}^{-2}$] & $\mathcal{U}\{[\rho_{P}^{\mathrm{min}}]^{2/3},[\rho_{P}^{\mathrm{max}}]^{2/3}\}$ \\
$i_{SB}$\,\,[rads] & $\mathcal{U}\{0,2\pi\}$ \\
$\Omega_{SB}$\,\,[rads] & $\mathcal{U}\{-\pi/2,\pi/2\}$ \\
$P_{SB}$\,\,$[\mathrm{days}]$ & $\mathcal{U}\{0.083,P_{B*}/\sqrt{3}\}$ \\
$\phi_{SB}$\,\,[rads] & $\mathcal{U}\{0,2\pi\}$ \\
$\sqrt{e_{SB}}\cos\omega_{SB}$ & $\mathcal{U}\{-1,1\}$ \\
$\sqrt{e_{SB}}\sin\omega_{SB}$ & $\mathcal{U}\{-1,1\}$ \\ [1ex]
\hline\hline %inserts single line
\end{tabular}
\label{tab:priors} % is used to refer this table in the text
\end{table}

\section{VETTING}
\label{sec:vetting}

There are two principal signals we are trying to detect: timing variations and
eclipse effects. Timing variations which could mimic moons can be induced by 
other perturbing bodies or even stellar activity. Similarly, light curve
distortions similar in morphology to mutual events can be induced by star
spot crossings. However, it should be noted that out-of-transit companion
eclipses are much harder to mimic.

Nevertheless, there is clearly a need to vet exomoon signals as being genuine
or not, in the same way candidate planets must be vetted. There are numerous 
tools at our disposal to aid in this procedure. Most generally, we follow the 
six detection criteria of \citet{thesis:2011} (see Chapter~1, \S4 for details):

\begin{itemize}
\item[{\textbf{C1}}] Statistically significant
\item[{\textbf{C2}}] Systematic errors dealt with
\item[{\textbf{C3}}] Physically plausible claim
\item[{\textbf{C4}}] Not a suspicious period(s)
\item[{\textbf{C5}}] Consistent instrumentation
\item[{\textbf{C6}}] Avoid large systematics (e.g. highly active stars)
\end{itemize}

Vetting is essentially the act of proposing alternative models which can perhaps
equally well, or even better, explain the observations. Thus we are considering
additional models $\mathcal{M}_2, \mathcal{M}_3, \mathcal{M}_4$, etc and
computing their evidences relative to that of $\mathcal{M}_1$, the 
planet-with-moon model. However, we note that in some cases, formally computing 
the evidence of an alternative model may not be necessary since it may 
immediately obvious that the observations cannot be caused by the alternative 
model in question.

\subsection{Weighing Tools}

The most powerful tool at our disposal for vetting will be the derived densities
from the light curve fits. As stated in \S\ref{sub:eclipsefeatures} \& 
\ref{sub:plancandidates}, exomoon systems allow one to measure the bulk density 
of the star, planet and moon all through photometry alone \citep{weighing:2010}. 
If one also has some radial velocity data, then the absolute dimensions of all 
three bodies can be determined using Kepler's Third Law.

If the moon is not genuine, but caused by some false-positive or even just 
residual noise in the data, then it is highly improbable that these values will 
come out to be remotely physical. Thus, by bounding the prior volume to only
physically plausible values (as discussed in \S\ref{sub:parameters}), we 
forcibly exclude the vast majority of false positives.

In most cases, we will not have a determination of the radial velocity 
semi-amplitude, $K_*$, at our disposal due to the faintness of the \emph{Kepler}
targets (although we may have an upper limit). Even without $K_*$, we still 
directly measure $\rho_*$, $\rho_P$ and $\rho_S$. Using the some reasonable 
estimate for $M_*$ and $R_*$, allows for the dimensions of all three bodies to 
be determined.

\subsection{Starspot Crossings}

Starspot crossings can mimic exomoon-like mutual events. There are several
ways in which they differ though. First of all, a starspot crossing tends
to be V-shaped (e.g. see \citealt{sanchis:2011}) due to the fact that the spots 
are typically greater than or approximately equal to the size of the transiting 
planet. In contrast, exomoon signatures should usually be flat-topped, but can 
in some more non-aligned geometries be V-shaped. Nevertheless, a flat-topped 
mutual event would be a strong indicator that the event was due to a moon rather 
than starspot crossing.

Secondly, starspots have chromatic variations whereas a mutual event should 
not. Thus there is the possibility of conducting follow-up observations from the 
ground to interrogate this hypothesis. Thirdly, starspots should track across
the stellar surface with a speed determined by the rotational period of the host
star, $P_{*,\mathrm{rot}}$. This term is usually determinable from the 
photometry alone, should the star be sufficiently active.

Since an auxiliary transit outside of the main transit event is much more
difficult to mimic through stellar activity, these events will undoubtedly
be more easily detected. These events tend to be associated with moons
on large separations such as a wide-captured retrograde moon and thus we
anticipate that HEK may have a detection bias in this regard.

However, starspots are unlikely to be both well-modelled by an exomoon-fit
and produce physically plausible densities for the star, planet and moon. This 
is because starspots tend to produce very strong eclipse features but only
weak timing artifacts. For example, HAT-P-11b transits a heavily spotted
star exhibiting eclipse features of up to 1-2\,mmag in depth 
\citep{sanchis:2011} but timing deviations of $\lesssim 30$\,seconds 
\citep{deming:2011}. The absence of significant timing variations would cause
the planet-with-moon fit to favour an implausibly low density moon. Thus,
the weighing tools will be a most common test for the presence of spots.

\subsection{Transit Timing Analysis}

Transit timing variations (TTVs) are induced in both resonant 
\citep{agol:2005,holman:2005} and non-resonant systems \citep{nesvorny:2010}.
One possible model to explain a candidate moon system would be a transiting
planet influenced by an unseen perturbing planet, $\mathcal{M}_2$. The
evidence of this model may be computed by allowing each transit epoch to have
its own unique time of transit minimum, $\tau_{B*,i}$, and proceeding as before
with the planet-only fit.

Aside from producing an evidence value more favourable than the planet-with-moon
fit, model $\mathcal{M}_2$ must correspond to a physically plausible scenario.
To investigate this, we will use the TTV-inversion code developed by 
\citet{nesvorny:2010}, \ttvim, to explore the range of plausible perturbers. If 
no plausible perturbers can fit the transit times, then model $\mathcal{M}_2$
will have a negligible prior probability i.e. 
$\mathrm{Pr}(\mathcal{M}_2)\rightarrow0$. Such an instance would thus favour the
planet-with-moon model over the perturber model. If the perturber is both
feasible and of comparable evidence to the moon-model, then one may search
the photometric time series for transits of the other planet directly.

The same process may be repeated for other sources of TTVs too, such as the 
model of a Trojan perturber \citep{ford:2007}, model $\mathcal{M}_3$. For such
scenarios, transits of the Trojan(s) can also be sought 
(e.g. \citealt{kippingbakos:2011a}).

\subsection{Radial Velocities}

Whenever possible, the HEK project will seek to obtain radial velocities of the
target systems to confirm both the planetary and lunar nature of the candidates.
Since TSA targets only sub-Jupiters on moderate-to-long periods, the expected RV
amplitudes will be typically $\lesssim 10$\,m/s. Detecting this signal would
allow us to both confirm the candidate and determine the absolute dimensions
of the system. However, excluding RV amplitudes above a certain threshold is
also useful in eliminating false positives.

Aside from dynamically weighing the system, we anticipate precise radial 
velocities will be very useful in locating the correct orbital period of the
moon. As discussed earlier and in \citet{kipping:2009a}, the timing signals
can be fitted by a forest of harmonic orbital periods due to the undersampled
nature of our observations. A very strong eclipse signal can remove this
degeneracy, as can enforcing coplanarity, but if either of these is not
possible, we are left with a forest of harmonics. Each period yields a
unique $\rho_P$ and thus unique light curve derived $M_P/M_*$. For example,
a synthetic example shown later in \S\ref{sub:NepM2far} finds two harmonic modes
with derived $\rho_P$ values differing by more than a factor of two. If
precision radial velocities can infer $M_P/M_*$ independently, then the correct
harmonic is empirically determined. Thus, systems with indistinguishable
harmonic modes will be prioritized for radial velocity follow-up.

\section{Worked Examples}
\label{sec:examples}

\subsection{Null Example: HZ Neptune around an M2 star}
\label{sub:nulleg}

As a null example, we use a synthetic planet-only light curve model from
\citet{luna:2011}. The synthetic data consists of six SC transits of a 
habitable-zone (HZ) Neptune around an M2-star, observed with \emph{Kepler}-class 
photometry (specifically we used Gaussian noise of 250\,ppm per minute). Details 
of the model can be found in \S4.6 of \citet{luna:2011}.

We fit the data with two models: $\mathcal{M}_0$, a planet-only model and
$\mathcal{M}_1$, a planet-with-moon model. In both cases we assume circular
orbits for simplicity. The adopted priors are the same as
that given in Table~\ref{tab:priors}. We used 
$[\rho_{P}^{\mathrm{min}}]^{2/3} = 18.4982$\,kg$^{2/3}$\,m$^{-2}$, corresponding
to the lower limit on the lowest density exoplanet presently known (WASP-17b;
\citealt{anderson:2010}), and 
$[\rho_{P}^{\mathrm{max}}]^{2/3} = 920.976$\,kg$^{2/3}$\,m$^{-2}$, corresponding
to a iron-rich Super-Earth \citep{valencia:2006}. For the stellar density,
we use the range provided in \citet{cox:2000} for main-sequence stars of
spectral type M5 to F0. $P_{B*}$ and $\tau_{B*}$ are given priors of $\pm 1$\,d
around the known solution. The solution for these terms is always easily
inferred from simple inspection of a photometric time series.

The Bayesian evidence for the two models was found to be
$\log(\mathcal{Z}_0) = 23377.36 \pm 0.21$ and 
$\log(\mathcal{Z}_1) = 23374.76 \pm 0.22$, thus giving
$|\Delta[\log \mathcal{B}]| = 2.60 \pm 0.30$ in favour of the planet-only model.
This occurs despite the fact the planet-with-moon model yields a lower $\chi^2$ 
of 8921.57, versus 8930.96 for the planet-only model, thus demonstrating the
built-in Occam's razor of Bayesian evidence.

\begin{figure*}
\begin{center}
\includegraphics[width=16.8 cm]{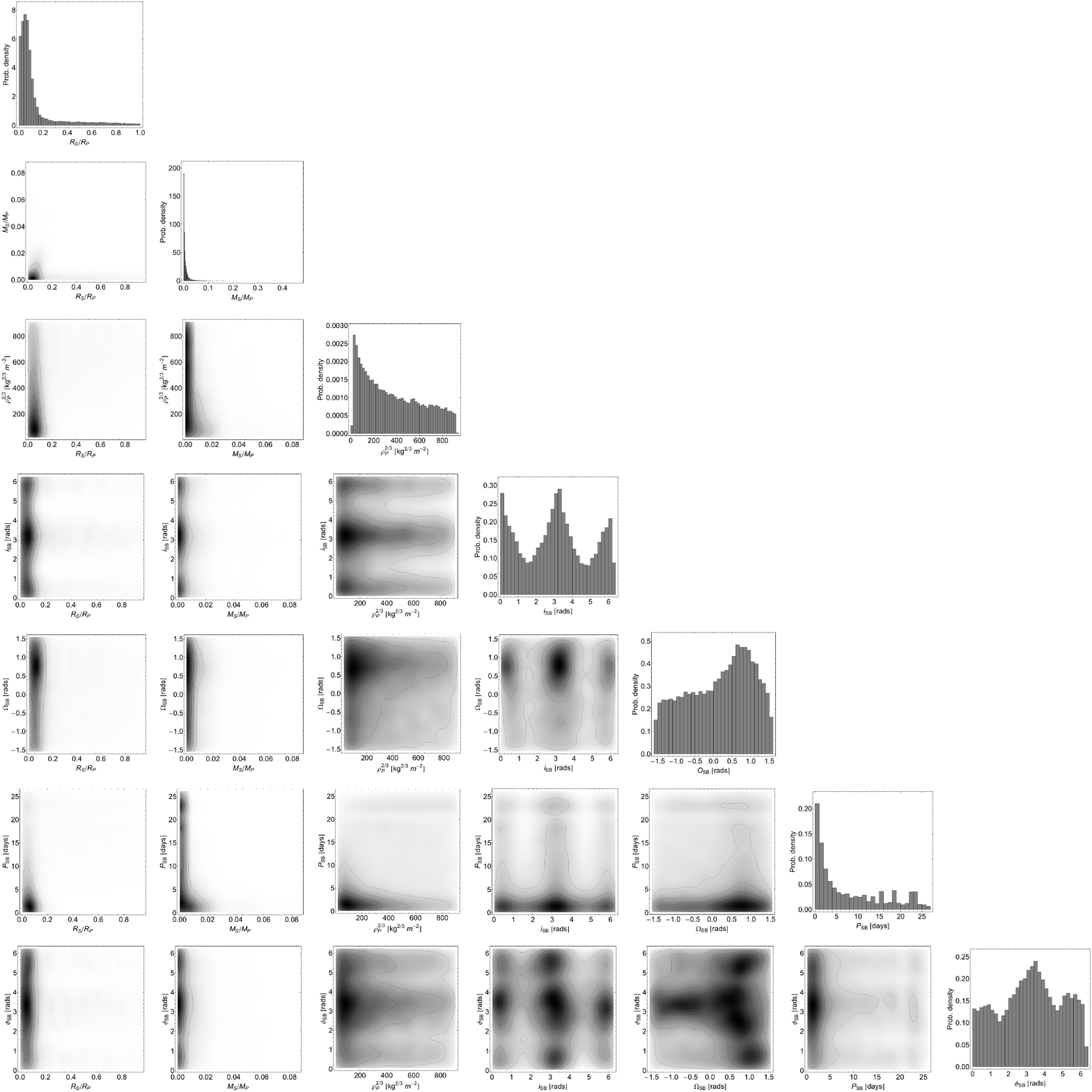}
\caption{\emph{
Marginalised posteriors from \multi\ when fitting a synthetic example of a 
planet-only data set using a planet-with-moon model from \luna. We here only 
show the exomoon-related parameters for brevity.
}} 
\label{fig:NepM2none_M1}
\end{center}
\end{figure*}

Despite the Bayesian evidence clearly favouring the planet-only model, as 
expected, the posteriors from the planet-with-moon model may be used to place 
upper limits on the allowed mass and radius of the exomoon. The posteriors, 
shown in Figure~\ref{fig:NepM2none_M1}, exclude $M_S/M_P > 0.018$ and 
$R_S/R_P > 0.44$ to 90\% confidence (the 4-$\sigma$ constraints are of limited 
use; $M_S/M_P > 0.369$ and $R_S/R_P > 0.999$).

%%% NULL HISTOS M0
\begin{figure*}
\begin{center}
\includegraphics[width=16.8 cm]{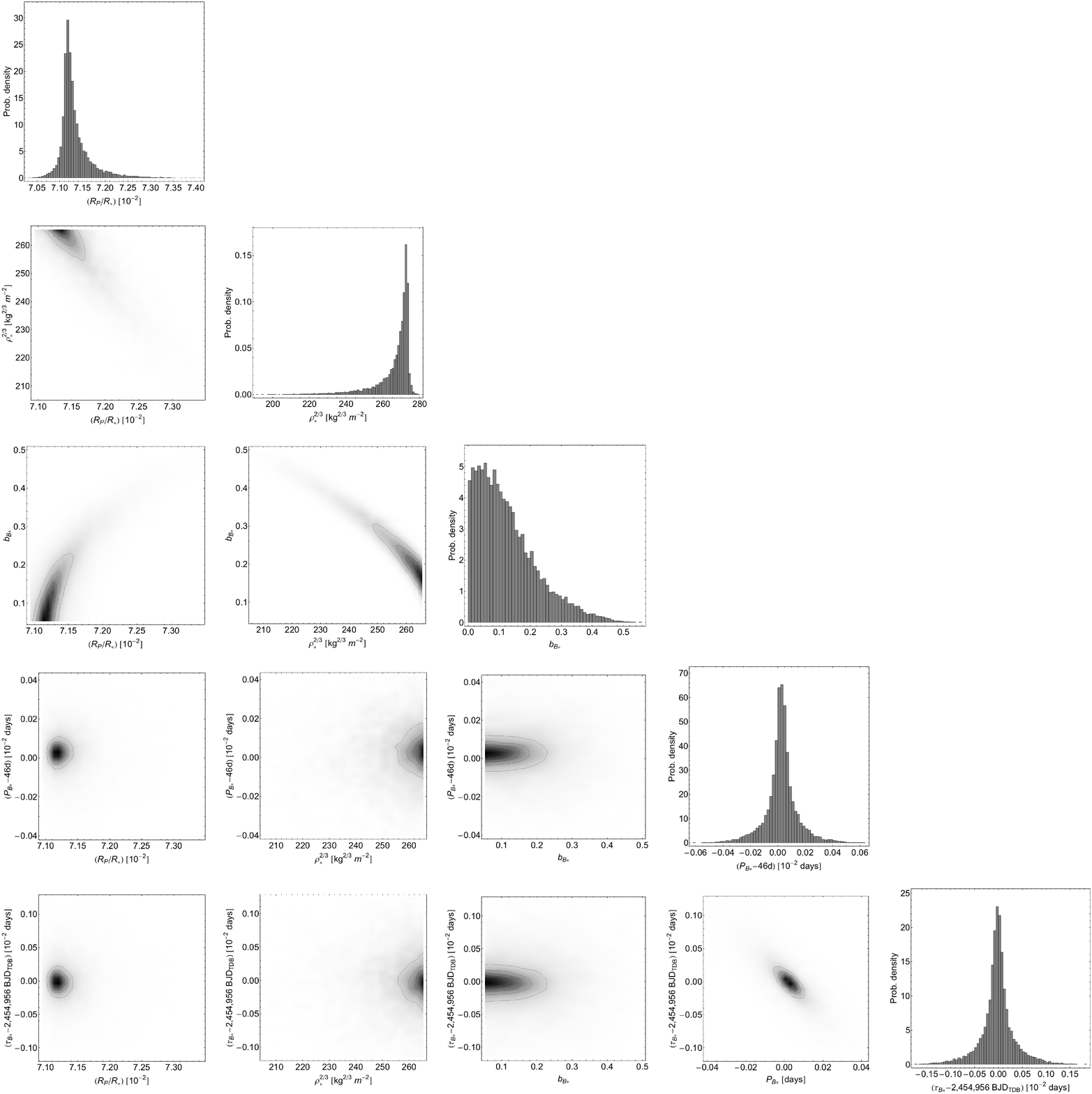}
\caption{\emph{
Marginalised posteriors from \multi\ when fitting a synthetic example of a 
planet-only data set using a planet-only model from \luna.
}} 
\label{fig:NepM2none_M0}
\end{center}
\end{figure*} 

%%% NULL HISTOS M1

\subsection{Moon Example: HZ Neptune around an M2 star with a Widely Separated 
Earth-like Moon}
\label{sub:NepM2far}

As a moon example, we use a synthetic planet-with-moon light curve model from
\citet{luna:2011}. The synthetic data consists of six SC transits of a 
habitable-zone (HZ) Neptune around an M2-star (as before), observed with 
\emph{Kepler}-class photometry (same noise as before). The moon is set to an 
Earth-mass/radius object at the edge of the Hill sphere on a retrograde orbit. 
Details of the model can be found in \S4.5 of \citet{luna:2011} and the 
de-noised model is shown in the left panel of Figure~\ref{fig:eclipses}.

\begin{figure*}
\begin{center}
\includegraphics[width=16.8 cm]{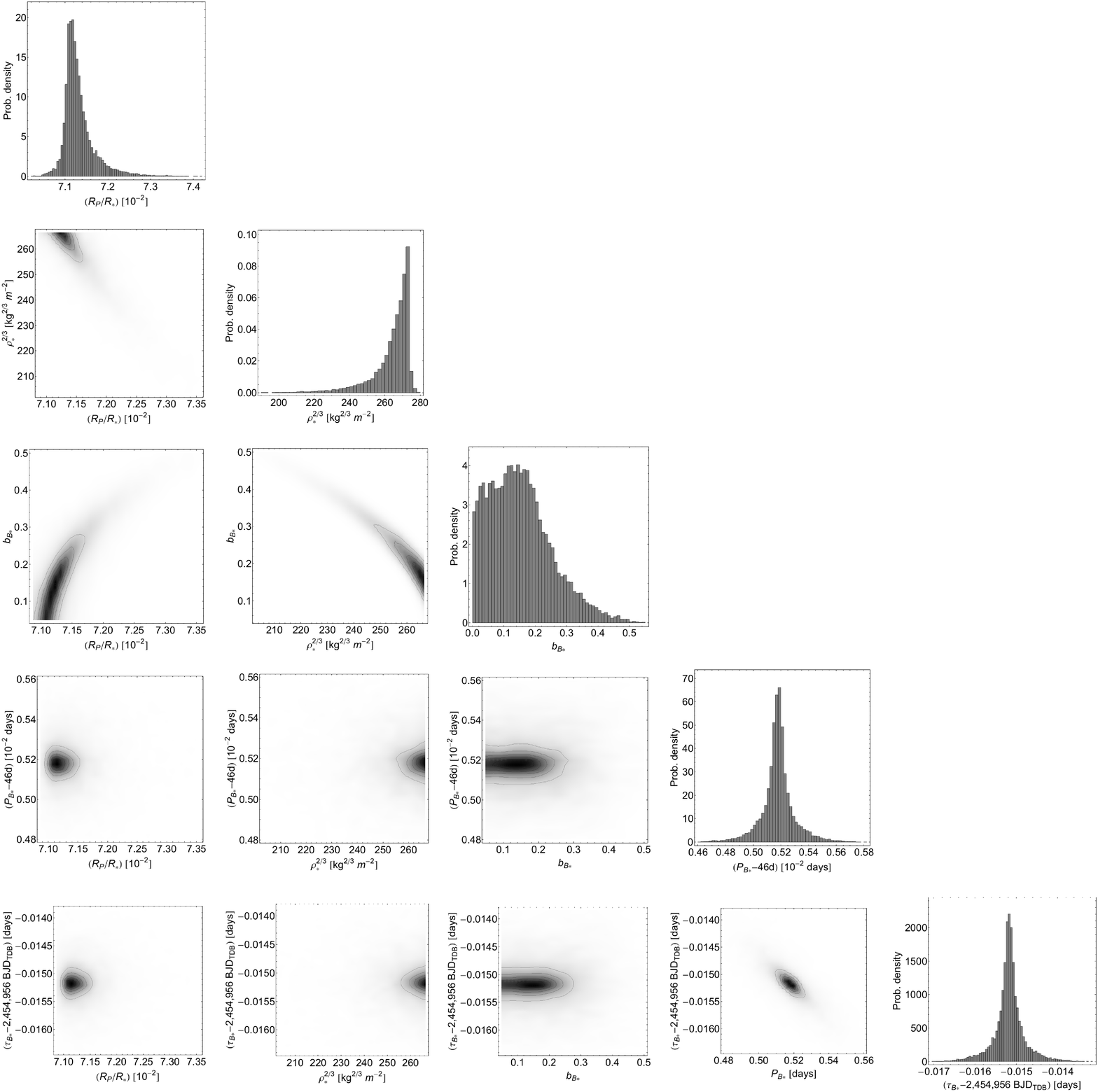}
\caption{\emph{
Marginalised posteriors from \multi\ when fitting a synthetic example of a 
planet-with-moon data set using a planet-only model from \luna.
}} 
\label{fig:NepM2far_M0}
\end{center}
\end{figure*}

\begin{figure*}
\begin{center}
\includegraphics[width=16.8 cm]{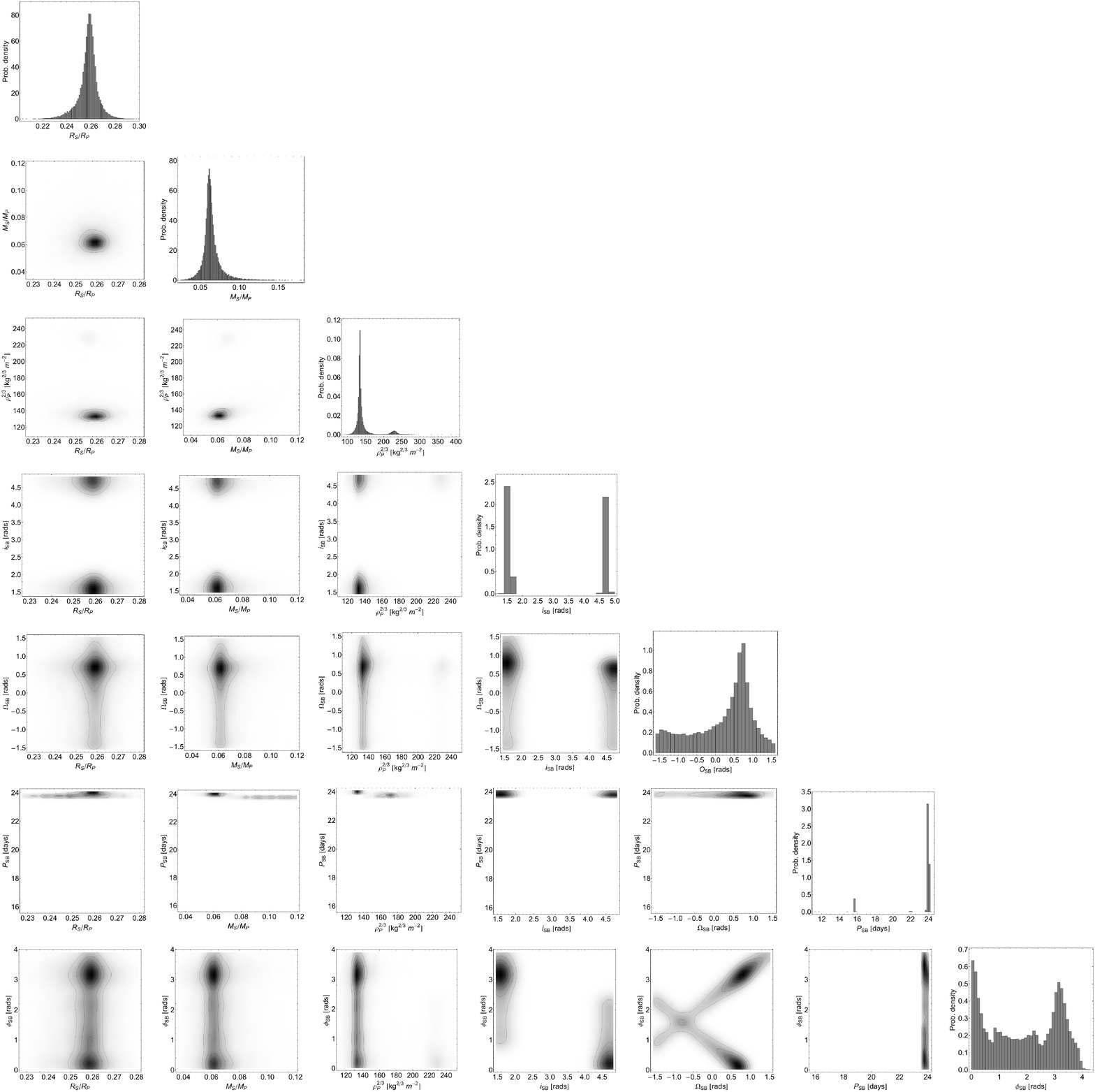}
\caption{\emph{
Marginalised posteriors from \multi\ when fitting a synthetic example of a 
planet-with-moon data set using a planet-with-moon model from \luna. We here 
only show the exomoon-related parameters for brevity.
}} 
\label{fig:NepM2far_M1}
\end{center}
\end{figure*}

Fitting the data using \multi\ with a planet-only model yields a Bayesian 
evidence of $\log\mathcal{Z}_0 = 22104.55 \pm 0.21$, whereas the 
planet-with-moon model reaps $\log\mathcal{Z}_1 = 23552.75 \pm 0.27$. The
corresponding posteriors of both models are shown in 
Figures~\ref{fig:NepM2far_M0}\&\ref{fig:NepM2far_M1}. The change in evidence 
corresponds to a $54$-$\sigma$ detection, which is close to the 50-$\sigma$
detection found using a simple F-test in \citet{luna:2011}. However, \multi\
reveals that three distinct modes exist in the data, which are reported
in Table~\ref{tab:NepM2far}. Modes 1 \& 3 correspond the correct exomoon
orbital period of $23.995$\,d, but mode 2 is located at $P_{SB} = 
15.775$\,d. This corresponds to a harmonic, as originally predicted in
\citet{kipping:2009a}. This is confirmed by evaluating the expected position
of the first harmonic using $[(1/P_{SB}) + (1/P_{B*})]^{-1} = 15.769$\,d.

Modes 1 and 3 both locate the true period and both exhibit a significantly 
improved evidence value. In fact, mode 2 is disfavoured at the $>$7-$\sigma$
level over modes 1 and 3. Between modes 1 and 3, the difference is that
mode 3 occurs at the correct retrograde solution whereas mode 1 is prograde.
The evidence difference between these modes is 
$|\Delta(\log\mathcal{Z}) = 1.2 \pm 0.6$ i.e. insignificant. In a blind
analysis, there would be no way of reliably distinguishing the correct solution 
based upon the available data, but more transits should help since the TDV-TIP
effect is asymmetric with respect to the orbital sense of motion.

Despite the presence of three modes, \multi\ provides an appropriately weighted
global posterior as well as the local posteriors. Using this weighted posterior
heavily deweights mode 2 and the leads to accurate values for the key terms
such as mass, radius and density (see last column of Table~\ref{tab:NepM2far}. 
Orbital angles such as inclination become unconstrained due to mixing the 
prograde and retrograde solutions together, but this is an accurate 
representation of our ignorance of this term.

\begin{table*}
\caption{\emph{Comparison of exomoon parameter estimates from three modes found 
in the \multi\ fits of synthetic data. Data generated for a Neptune with a 
distant moon around an M2 star. The global evidence of the planet-with-moon 
model (all three modes) is $\log\mathcal = $. Mode 3 is the 
most accurate mode but the local evidence values between modes 1 and 3 are 
sufficiently close that distinguishing these modes blindly would not be possible 
(mode 2, however, can be discounted).}} % title of Table
\centering % used for centering table
\begin{tabular}{c c c c c c} % centered columns (3 columns)
\hline\hline %inserts double horizontal lines
\textbf{Parameter} & \textbf{Truth} & \textbf{Mode 1} & \textbf{Mode 2} & \textbf{Mode 3} & \textbf{Global} \\ [0.5ex] % inserts table
%heading
\hline
$\log\mathcal{Z}$ & - & $23552.49 \pm 0.36$ & $23523.81 \pm 0.99$ & $23551.27 \pm 0.45$ & $23552.75 \pm 0.27$ \\
\hline
\emph{Moon params.} & & & & & \\
\hline % inserts single horizontal line
$R_S/R_P$ & $0.2570$ & $0.2587_{-0.0069}^{+0.0053}$ & $0.2559_{-0.0065}^{+0.0051}$ & $0.2587_{-0.0070}^{+0.0052}$ & $0.2585_{-0.0069}^{+0.0053}$ \\
$M_S/M_P$ & $0.0583$ & $0.0620_{-0.0058}^{+0.0086}$ & $0.0672_{-0.0073}^{+0.0072}$ & $0.0622_{-0.055}^{+0.0096}$ & $0.0624_{-0.0059}^{+0.0092}$ \\ 
$[\rho_{P}]^{2/3}$ [kg$^{2/3}$\,m$^{-2}$] & $139.0$ & $134.5_{-4.7}^{+5.8}$ & $229.4_{-8.9}^{+10.2}$ & $134.7_{-4.4}^{+6.6}$ & $134.9_{-4.5}^{+11.0}$ \\
$i_{SB}$ [$^{\circ}$] & $267.06$ & $90.1_{-1.5}^{+1.4}$ & $270.20_{-0.70}^{+1.20}$ & $270.1_{-1.2}^{+1.3}$ & $90_{-3}^{+180}$ \\
$\Omega_{SB}$ [$^{\circ}$] & $5$ & $37_{-72}^{+24}$ & $25_{-71}^{+18}$ & $16_{-69}^{+24}$ & $28_{-73}^{+23}$ \\
$P_{SB}$ [days] & $23.995$ & $23.990_{-0.047}^{+0.020}$ & $15.755_{-0.014}^{+0.010}$ & $23.989_{-0.053}^{+0.019}$ & $23.987_{-0.081}^{+0.021}$ \\
$\phi_{SB}$ [$^{\circ}$] & $40$ & $173_{-72}^{+25}$ & $23_{-18}^{+70}$ & $31_{-24}^{+69}$ & $112_{-95}^{+75}$ \\ [1ex]
\hline\hline %inserts single line
\end{tabular}
\label{tab:NepM2far} % is used to refer this table in the text
\end{table*}

%%% POS HISTOS M0

%%% POS HISTOS M1

\section{SUMMARY}
\label{sec:summary}

In this work, we have presented a description of the objectives and methods of 
our new observational project, ``The Hunt for Exomoons with Kepler'' (HEK). The
HEK project will seek to infer the presence of extrasolar moons around 
transiting exoplanet candidates observed by the \emph{Kepler Mission}. In cases
of null detections, upper limits will be reported and the full set of
parameter posteriors will be made available on the project website 
(www.cfa.harvard.edu/HEK/). These statistics may be used to deduce the
frequency of large moons around viable exoplanet hosts, $\eta_{\leftmoon}$.

We have described, in \S\ref{sec:targets}, how the list of all Kepler Objects of 
Interest (KOIs) will be distilled into a subset of the most promising 
candidates, for the purposes of exomoon detection, via a three-prong target
selection (TS) strategy. This includes visual identification (TSV), automatic
filtering (TSA) and targets-of-opportunity (TSO) with a final stage of
target prioritization (TSP).

Selected targets will be interrogated for evidence of an exomoon by comparing
the Bayesian evidence, $\mathcal{Z}$ of a planet-only model and a 
planet-with-moon model (see discussion in \S\ref{sub:bayesian}). In addition to 
presenting a $>4$-$\sigma$ preference for the planet-with-moon model, putative 
candidates must demonstrate physically plausible solutions.

In fitting the data, we require both a forward-model and a fitting algorithm.
The former task is handled by the \luna\ algorithm, developed by 
\citet{luna:2011}, which is designed to analytically model the transit
light curve of a planet-with-moon system including limb darkening, dynamical
motion and mutual events. Due to the highly complex parameter space, featuring
multiple modes due to aliased harmonic power, and the need for an highly
efficient computation of the Bayesian evidence, a sophisticated fitting
algorithm is required. To this end, we have presented the application of \multi\
\citep{feroz:2007} with \luna. \multi\ is a multimodal nested sampling algorithm
widely used in the cosmology and particle physics communities. 
\S\ref{sub:fitting} describes our implementation and the choice of priors,
parameterization and likelihood function.

Strategies for vetting potential candidates are discussed in \S\ref{sec:vetting}
before we present two examples of \luna+\multi\ on synthetic data in
\S\ref{sec:examples}. The example fits demonstrate not only the multimodal 
nature of looking for exomoons but also how Bayesian evidence may be used to 
detect such systems.

We are currently analyzing a subset of preferred candidates and will be
reporting on these findings in the near-future (Kipping et al. 2012, in prep.).
As the HEK project progresses, we hope to answer the question as to whether 
large moons, possibly even Earth-like habitable moons, are common in the
Galaxy or not. Enabled by the exquisite photometry of \emph{Kepler}, exomoons
may soon move from theoretical musings to objects of empricial investigation.

% #####################################################################
%% Acknowledgements
\acknowledgements
\section*{Acknowledgements}

We offer our thanks and praise to the extraordinary scientists, engineers
and individuals who have made the \emph{Kepler Mission} possible. Without
their continued efforts and contribution, our project would not be possible.
We are very grateful to the PlanetHunters.org community for their interest
in extrasolar moons and their assistance in identifying interesting candidate
signals. Special thanks to Farhan Feroz, Mike Hobson and Sree Balan for 
discussions regarding the application of {\sc MultiNest}. Thanks to the
anonymous referee for their insightful and helpful comments which led to
an overall improved manuscript.

%% EOF Acknowledgements

% #####################################################################
%% Bibliography
% #####################################################################
%% Bibliography

\clearpage

\section*{Appendix}

\begin{center}
\begin{longtable}{ll}
\caption[List of important acronyms used in this paper]
{\emph{List of important acromyms used in this paper.}} \\ % title of Table
%\centering % used for centering table
%\begin{tabular}{l l} % centered columns (5 columns)
\hline\hline %inserts double horizontal lines
\textbf{Acromyn} & \textbf{Definition} \\ [0.5ex] % inserts table
%heading
\hline % inserts single horizontal line
HEK & Hunt for Exomoons with \emph{Kepler} project \\
SC & short-cadence \\
LC & long-cadence \\
KIC & Kepler Input Catalogue \\
TTV & Transit timing variations \\
TDV & Transit duration variations \\
TDV-V & Velocity induced TDV \\
TDV-TIP & Transit impact parameter induced TDV \\
TS & Target selection \\
TSA & Automatic target selection \\
TSV & Visual target selection \\
TSP & Prioritization target selection \\ [1ex]
\hline\hline %inserts single line
%\end{tabular}
\label{tab:acros} % is used to refer this table in the text
\end{longtable}
\end{center}

\begin{center}
\begin{longtable}{ll}
\caption[List of important parameters used in this paper]
{\emph{List of important parameters used in this paper.}} \\ % title of Table
%\centering % used for centering table
%\begin{tabular}{l l} % centered columns (5 columns)
\hline\hline %inserts double horizontal lines
\textbf{Parameter} & \textbf{Definition} \\ [0.5ex] % inserts table
%heading
\hline % inserts single horizontal line
%%%%% GENERAL CARTESIANS
%$S'$ & Sky-projected separation between two objects \\
%$S$ & Sky-projected separation between two objects, in units of $R_*$ \\
%$X'$ & Sky-projected $X$-component position of an object \\
%$X$ & Sky-projected $X$-component position of an object, in units of $R_*$ \\
%$Y'$ & Sky-projected $Y$-component position of an object \\
%$Y$ & Sky-projected $Y$-component position of an object, in units of $R_*$ \\
%$Z'$ & Sky-projected $Z$-component position of an object \\
%$Z$ & Sky-projected $Z$-component position of an object, in units of $R_*$ \\
%%%%% P* CARTESIANS 
%$S_{P*}'$ & Sky-projected planet-star separation \\
%$S_{P*}$ & Sky-projected planet-star separation, in units of $R_*$ \\
$S_{B*}$ & Sky-projected separation between the planet-moon barycentre and the host star, in units of $R_*$ \\
$\nu_{SB}$ & True anomaly of the satellite around the planet-moon barycentre \\
%$E_S$ & Eccentric anomaly of the satellite around the host planet\\
%$M_S$ & Mean anomaly of the satellite around the host planet \\
%%%%% ORBITAL ELEMENTS
%$e$ & Orbital eccentricity of an object around its primary \\
$e_{B*}$ & Orbital eccentricity of the barycentre of the planet (+ any satellites) around the host star \\
$e_{SB}$ & Orbital eccentricity of the satellite around the planet-moon barycentre \\
%$\omega$ & Argument of periapsis of an object around its primary \\
$\omega_{B*}$ & Argument of periapsis of the barycentre of the planet (+ any satellites) around the host star \\
$\omega_{SB}$ & Argument of periapsis of the satellite around the planet-moon barycentre \\
%$k_P$ & Lagrangian eccentricity parameter $k$ for the planet; $k_P = e_P \cos \omega_P$ \\
%$h_P$ & Lagrangian eccentricity parameter $h$ for the planet; $h_P = e_P \sin \omega_P$ \\
%$k_P$ & Lagrangian eccentricity parameter $k$ for the satellite; $k_S = e_S \cos \omega_S$ \\
%$k_P$ & Lagrangian eccentricity parameter $h$ for the satellite; $h_S = e_S \sin \omega_S$ \\
%$i$ & Orbital inclination of an object around its primary \\
$i_{B*}$ & Orbital inclination of the barycentre of the planet (+ any satellites) around the host star \\
$i_{SB}$ & Orbital inclination of the satellite around the planet-moon barycentre \\
%$\Omega$ & Longitude of the ascending node of an object around its primary \\
%$\Omega_P$ & Longitude of the ascending node of a planet, relative to the sky-plane \\
$\Omega_{SB}$ & Longitude of the ascending node of a satellite, relative to the orbital plane of the planet \\
%$\varpi_S$ & Longitude of the periapsis of a satellite, defined as $\varpi_S = \omega_P + \Omega_S$ \\
%%%%% PHYSICAL PARAMETERS
%$R$ & Radius of an object \\
$R_*$ & Radius of the host star \\
$R_P$ & Radius of the planet \\ 
$R_S$ & Radius of the satellite \\ 
$a$ & Semi-major axis of an object around its primary \\
%$a_{P*} = a_P$ & Semi-major axis of the planet around the host star \\
$a_{B*}$ & Semi-major axis of the planet-moon barycentre around the host star \\
%$a_{SP} = a_S$ & Semi-major axis of the satellite around the host planet \\
$a_{SB}$ & Semi-major axis of the satellite around the planet-moon barycentre \\
%$a_{PB}$ & Semi-major axis of the planet around the planet-moon barycentre \\
%$P$ & Orbital period of an object around its primary \\
$P_{B*}$ & Orbital period of the barycentre of the planet (+ any satellites) around the host star \\
$P_{SB}$ & Orbital period of the satellite around the planet-moon barycentre \\
%$M$ & Mass of an object \\
$M_*$ & Mass of the star \\
$M_P$ & Mass of the planet \\
$M_S$ & Mass of the satellite \\
%$\rho$ & Mean density of an object \\
$\rho_*$ & Mean density of the star \\
$\rho_P$ & Mean density of the planet \\
$\rho_S$ & Mean density of the satellite \\
%$r$ & Separation of an object from its primary \\
%$r_{P*} = r_P$ & Separation of the planet from the host star \\
%$r_{B*}$ & Separation of the planet-moon barycentre from the host star \\
%$r_{SP} = r_S$ & Separation of the satellite from the host planet \\
%$r_{SB}$ & Separation of the satellite from the planet-moon barycentre \\
%$r_{PB}$ & Separation of the satellite from the planet-moon barycentre \\
%$\varrho$ & Separation of an object from its primary, in units of $a$ \\
%$\varrho_{P*} = r_P$ & Separation of the planet from the host star, in units of $a_{P*}=a_P$ \\
%$\varrho_{B*}$ & Separation of the planet-moon barycentre from the host star, in units of $a_{P*}=a_P$ \\
%$\varrho_{SP} = r_S$ & Separation of the satellite from the host planet, in units of $a_{SP}=a_S$ \\
%$\varrho_{SB}$ & Separation of the satellite from the planet-moon barycentre, in units of $a_{SB}$ \\
%$\varrho_{PB}$ & Separation of the satellite from the planet-moon barycentre, in units of $a_{PB}$ \\
%%% TRANSIT TERMS
$p$ & Ratio of the planet's radius to the stellar radius ($R_P/R_*$) \\
$s$ & Ratio of the satellite's radius to the stellar radius ($R_S/R_*$) \\
$\delta$ & Defined as $p^2$ \\
$\tau_{B*}$ & Instant when $\mathrm{d}S_{B*}/\mathrm{d}t = 0$ near inferior conjunction \\
%$\tau_{O}$ & Instant when $\mathrm{d}S_{P*}/\mathrm{d}t = 0$ near superior conjunction \\
$b_{B*}$ & Impact parameter of the barycentre of the planet (+ any satellites) \\
%$b_{SB}$ & Impact parameter of the satellite, defined as $r_{SB}\cos i_S/R_P$ \\
%$t_{I}$ & Instant when $S_{B*} = 1+p$ and $\dot{S_{B*}} < 0$ \\
%$t_{II}$ & Instant when $S_{B*} = 1-p$ and $\dot{S_{B*}} < 0$ \\
%$t_{III}$ & Instant when $S_{B*} = 1-p$ and $\dot{S_{B*}} > 0$ \\
%$t_{IV}$ & Instant when $S_{B*} = 1+p$ and $\dot{S_{B*}} > 0$ \\
%$T_{xy}$ & Time for a planet to move between contact points x and y \\
$\tilde{T}$ & Time between the planet's centre crossing the stellar limb to exiting under the same condition \\
%%% TTV/TDV TERMS
$\delta_{\mathrm{TTV}}$ & RMS amplitude of TTV signal \\
$\delta_{\mathrm{TDV-V}}$ & RMS amplitude of TDV-V signal \\
$\delta_{\mathrm{TDV-TIP}}$ & RMS amplitude of TDV-TIP signal \\
%$\delta_{\mathrm{TDV}}$ & RMS amplitude of total TDV signal \\
$\Lambda_{\mathrm{TTV}}$ & Waveform of TTV signal \\
$\Lambda_{\mathrm{TDV-V}}$ & Waveform of TDV-V signal \\
$\Lambda_{\mathrm{TDV-TIP}}$ & Waveform of TDV-TIP signal \\
%$\Lambda_{\mathrm{TDV}}$ & Waveform of total TDV signal \\
$\Phi_{\mathrm{TTV}}$ & Enhancement factor of TTV signal \\
$\Phi_{\mathrm{TDV-V}}$ & Enhancement factor of TDV-V signal \\
$\Phi_{\mathrm{TDV-TIP}}$ & Enhancement factor of TDV-TIP signal \\
%$\Phi_{\mathrm{TDV}}$ & Enhancement factor of total TDV signal \\
%$\eta$ & Equal to $\delta_{\mathrm{TDV}}/\delta_{\mathrm{TTV}}$ \\ [1ex]
%%% TIDAL BITS
$Q_P$ & Tidal quality factor of the planet \\
$k_{2p}$ & Love number of the planet \\
$\mathbb{T}$ & Lifetime of the moon \\ [1ex]
\hline\hline %inserts single line
%\end{tabular}
\label{tab:parameters} % is used to refer this table in the text
\end{longtable}
\end{center}


\begin{thebibliography}{99}
\bibitem[\protect\citeauthoryear{Abdussalam et al.}{2010}]{abdussalam:2010} 
Abdussalam, S. S., Allanach, B. C., Quevedo, F., Feroz, F. \& Hobson, M.,
2010, Phys. Rev., D81, 035017
\bibitem[\protect\citeauthoryear{Agnor \& Hamilton}{2006}]{agnor:2006} Agnor, 
C. B. \& Hamilton, D. P., 2006, Nature, 441, 192
\bibitem[\protect\citeauthoryear{Agol et al.}{2005}]{agol:2005} Agol, E., 
Steffen, J., Sari, R. \& Clarkson, W., 2005, MNRAS, 359, 567
\bibitem[\protect\citeauthoryear{Anderson et al.}{2010}]{anderson:2010} 
Anderson, D. R. et al., 2010, ApJ, 709, 159
\bibitem[\protect\citeauthoryear{Bakos et al.}{2009}]{bakos:2009} Bakos, G. A.
et al., 2009, ApJ, 710, 1724
\bibitem[\protect\citeauthoryear{Barnes \& O'Brien}{2002}]{barnes:2002} 
Barnes, J. W. \& O'Brien, D. P., 2002, ApJ, 575, 1087
\bibitem[\protect\citeauthoryear{Batalha et al.}{2012}]{batalha:2011}
Batalha, N. et al., 2012, in prep.
\bibitem[\protect\citeauthoryear{Borucki et al.}{2011}]{borucki:2011} 
Borucki, W. J. et al., 2011, ApJ, 736, 19
\bibitem[\protect\citeauthoryear{Ter Braak}{2006}]{braak:2006} 
Ter Braak, C. J. F., 2006, Stat. Comput. 16, 239
\bibitem[\protect\citeauthoryear{Brown et al.}{2001}]{brown:2001} 
Brown, T. M., Charbonneau, D., Gilliland, R. L., Noyes, R. W. \& Burrows, A.,
2001, ApJ, 552, 699
\bibitem[\protect\citeauthoryear{Brown et al.}{2011}]{brown:2011} 
Brown, T. M., Latham, D. W., Everett, M. E. \& Esquerdo, G. A., 2011, 
AJ, 142, 112
\bibitem[\protect\citeauthoryear{Canup \& Ward}{2006}]{canup:2006} 
Canup, R. M. \& Ward, W. R., 2006, Nature, 441, 834
\bibitem[\protect\citeauthoryear{Carter et al.}{2008}]{carter:2008} 
Carter, J. A., Yee, J. C., Eastman, J., Gaudi, B. S. \& Winn, J. N., 2008, ApJ, 
689, 499
\bibitem[\protect\citeauthoryear{Carter et al.}{2009}]{carter:2009} 
Carter, J. A., Winn, J. N., Gilliland, R. \& Holman, M. J., 2009, ApJ, 
696, 241
\bibitem[\protect\citeauthoryear{Cassidy et al.}{2009}]{cassidy:2009}
Cassidy, T. A., Mendez, R., Arras, P., Johnson, R. E. \& Skrutskie, M. F.,
2009, ApJ, 704, 1341
\bibitem[\protect\citeauthoryear{Castelli \& Kurucz}{2004}]{castelli:2004} 
Castelli, F. \& Kurucz, R. L., 2004, astro-ph:0405087
\bibitem[\protect\citeauthoryear{Cochran et al.}{2011}]{cochran:2011} 
Cochran, W. D. et al., 2011, ApJS, 197, 7
\bibitem[\protect\citeauthoryear{Claret}{2000}]{claret:2000} Claret A., 2000, 
A\&A, 363, 1081
\bibitem[\protect\citeauthoryear{Cox}{2000}]{cox:2000} Cox, A. N. (ed.), Allen's 
Astrophysical Quantities (4th edition) (Springer, Heidelberg), 2000.
\bibitem[\protect\citeauthoryear{D'Angelo et al.}{2010}]{angelo:2010} 
D'Angelo G., Durisen R. H. \& Lissauer J. J., 2010, Exoplanets, ed. S. Seager,
University of Arizona Press, p. 319
\bibitem[\protect\citeauthoryear{Deeg}{2009}]{deeg:2009} 
Deeg, H. 2009, IAU Symp., 253, 388
\bibitem[\protect\citeauthoryear{Deleuil et al.}{2008}]{deleuil:2008} 
Deleuil, M. et al., 2008, A\&A, 491, 889
\bibitem[\protect\citeauthoryear{Deming et al.}{2011}]{deming:2011} 
Deming, D. et al., 2011, ApJ, 740, 33
\bibitem[\protect\citeauthoryear{Domingos et al.}{2006}]{domingos:2006} 
Domingos, R. C., Winter, O. C. \& Yokoyama, T., 2006, MNRAS, 373, 1227
\bibitem[\protect\citeauthoryear{Eberle et al.}{2010}]{eberle:2011} 
Eberle, J. Cuntz, M., Quarles, B. \& Musielak, Z. E., 2011, Int. Jour. 
Astrobiology, 10, 325
\bibitem[\protect\citeauthoryear{Elser et al.}{2011}]{elser:2011} 
Elser, S., Moore, B., Stadel, J. \& Morishima, R., 2011, Icarus, 214, 357
\bibitem[\protect\citeauthoryear{Feroz et al.}{2007}]{feroz:2007} 
Feroz, F. \& Hobson, M. P., 2007, MNRAS, 384, 449
\bibitem[\protect\citeauthoryear{Feroz et al.}{2009a}]{feroz:2008} 
Feroz, F., Hobson, M. P. \& Bridges, M., 2009a, MNRAS, 398, 1601
\bibitem[\protect\citeauthoryear{Feroz et al.}{2009b}]{feroz:2009} 
Feroz, F., Gair, J. R., Hobson, M. P. \& Porter, E. K., 2009b, CQG, 26, 215003
\bibitem[\protect\citeauthoryear{Feroz et al.}{2011}]{feroz:2011} 
Feroz, F., Balan, S. T. \& Hobson, M. P., 2011, MNRAS, 416, L104
\bibitem[\protect\citeauthoryear{Fewell}{2006}]{fewell:2006} 
Fewell, M., ``Area of Common Overlap of Three Circles'', Tech. Rep. 
DSTO-TN-0722, 2006. [Online]. Available: http://hdl.handle.net/1947/4551
\bibitem[\protect\citeauthoryear{Fischer et al.}{2012}]{fischer:2012} 
Fischer, D. A. et al., 2012, MNRAS, 419, 2900
\bibitem[\protect\citeauthoryear{Ford \& Holman}{2007}]{ford:2007} Ford, E. B. 
\& Holman, M. J., 2007, ApJ, 664, 51
\bibitem[\protect\citeauthoryear{Fossey et al.}{2012}]{fossey:2012} 
Fossey, S. J., Kipping, D. M. et al., 2012, in prep.
\bibitem[\protect\citeauthoryear{Geyer}{1991}]{geyer:1991}
Geyer, C. J., 1991, in ``Computing Science and Statistics: 23rd symposium on the 
interface'', American Statistical Association, New York, p. 156
\bibitem[\protect\citeauthoryear{Gillon et al.}{2011}]{gillon:2011}
Gillon, M. et al., 2012, A\&A, 539, 28
\bibitem[\protect\citeauthoryear{Girardi et al.}{2000}]{girardi:2000}
Girardi, L., Bressan, A., Bertelli, G., \& Chiosi, C., 2000, A\&AS,
141, 371
\bibitem[\protect\citeauthoryear{Goldreich \& Soter}{1966}]{goldreich:1966}
Goldreich, P. \& Soter, S. 1966, Icarus, 5, 375
\bibitem[\protect\citeauthoryear{Gregory}{2009}]{gregory:2009}
Gregory, P. C., 2009, in ``Bayesian Inference and Maximum Entropy Methods in 
Science and Engineering: 27th International Workshop'', Saratoga Springs, eds. 
K. H. Knuth, A. Caticha, J. L. Center, A,G ̇ iffin, C. C. Rodrguez, AIP 
Conference Proceedings, 954, 307
\bibitem[\protect\citeauthoryear{Heller et al.}{2011}]{heller:2011} 
Heller, R. Leconte, J. \& Barnes, R., 2011, A\&A, 528, 27
\bibitem[\protect\citeauthoryear{Henry et al.}{2011}]{henry:2011} 
Henry, G. W., Howard, A. W., Marcy, G. W., Fischer, D. A. \& Johnson, J. A.,
2011, ApJ, submitted (astro-ph:1109.2549)
\bibitem[\protect\citeauthoryear{Holman \& Murray}{2005}]{holman:2005} 
Holman, M. J. \& Murray, N. W., 2005, Science, 307, 1288
\bibitem[\protect\citeauthoryear{Hubbard}{1984}]{hubbard:1984} 
Hubard W. B., 1984, Planetary Interiors (New York: Van Nostrand
Reinhold)
\bibitem[\protect\citeauthoryear{Kipping}{2009a}]{kipping:2009a} Kipping, D. M., 
2009a, MNRAS, 392, 181
\bibitem[\protect\citeauthoryear{Kipping}{2009b}]{kipping:2009b} Kipping, D. M., 
2009b, MNRAS, 396, 1797
\bibitem[\protect\citeauthoryear{Kipping et al.}{2009}]{kipping:2009} 
Kipping, D. M., Fossey, S. J. \& Campanella, G., 2009, MNRAS, 400, 398
\bibitem[\protect\citeauthoryear{Kipping}{2010a}]{investigations:2010} 
Kipping, D. M., 2010a, MNRAS, 407, 301
\bibitem[\protect\citeauthoryear{Kipping}{2010b}]{binning:2010} 
Kipping, D. M., 2010b, MNRAS, 408, 1758
\bibitem[\protect\citeauthoryear{Kipping}{2010c}]{weighing:2010} 
Kipping, D. M., 2010c, MNRAS, 409, L119
\bibitem[\protect\citeauthoryear{Kipping \& Bakos}{2011a}]{kippingbakos:2011a} 
Kipping, D. M. \& Bakos, G. A., 2011a, ApJ, 730, 50
\bibitem[\protect\citeauthoryear{Kipping \& Bakos}{2011b}]{kippingbakos:2011b} 
Kipping, D. M. \& Bakos, G. A., 2011b, ApJ, 733, 36
\bibitem[\protect\citeauthoryear{Kipping \& Tinetti}{2010}]{kiptin:2010} 
Kipping, D. M. \& Tinetti, G., 2010, MNRAS, 407, 2589
\bibitem[\protect\citeauthoryear{Kipping}{2011a}]{luna:2011} 
Kipping, D. M., 2011a, MNRAS, 416, 689
\bibitem[\protect\citeauthoryear{Kipping}{2011b}]{thesis:2011} 
Kipping, D. M., 2011b, PhD thesis, University College London (astro-ph:1105.3189)
\bibitem[\protect\citeauthoryear{Kipping et al.}{2012}]{map:2012} 
Kipping, D. M., Dunn, W., Jasinski, J. \& Manthri, V. M., 2012, MNRAS, accepted
(astro-ph:1112.2700)
\bibitem[\protect\citeauthoryear{Kundurthy et al.}{2011}]{kundurthy:2011} 
Kundurthy, P., Agol, E., Becker, A. C., Barnes, R., Williams, B \& Mukadam, A.,
2011, ApJ, 731, 123
\bibitem[\protect\citeauthoryear{Lewis}{2011}]{lewis:2011} 
Lewis, K. M. 2011, PhD thesis, Monash Univ., Australia (astro-ph:1109.5332) 
\bibitem[\protect\citeauthoryear{Liddle et al.}{2007}]{liddle:2007} 
Liddle, A. R., 2007, MNRAS, 377, L74
\bibitem[\protect\citeauthoryear{Lintott et al.}{2012}]{lintott:2012} 
Lintott, C. et al., 2012, AJ, submitted (astro-ph:1202.6007)
\bibitem[\protect\citeauthoryear{Mandel \& Agol}{2002}]{mandel:2002} Mandel, K. 
\& Agol, E., 2002, ApJ, 580, 171
\bibitem[\protect\citeauthoryear{Mislis et al.}{2011}]{mislis:2012} Mislis, D.,
Heller, R., Schmitt, J. H. M. M. \& Hodgkin, S. 2011, astro-ph:1112.2008
\bibitem[\protect\citeauthoryear{Muirhead et al.}{2011}]{muirhead:2011} 
Muirhead, P., Hamren, K., Schlawin, E., Rojas-Ayala, B., Covey, K. R. \& Lloyd, 
J. P., 2011, ApJL, submitted (astro-ph:1109.1819)
\bibitem[\protect\citeauthoryear{Mukherjee et al.}{2006}]{mukherjee:2006}
Mukherjee P., Parkinson D. \& Liddle A. R., 2006, ApJ, 638, L51
\bibitem[\protect\citeauthoryear{Namouni}{2010}]{namouni:2010} Namouni, F.,
2010, ApJ, 719, 145
\bibitem[\protect\citeauthoryear{Nesvorn\'{y} \& Beaug\'{e}}{2010}]{nesvorny:2010} 
Nesvorn\'{y}, D. \& Beaug\'{e}, C., 2010, ApJ, 709, 44
\bibitem[\protect\citeauthoryear{\'{O} Ruanaidh \& Fitzgerald}{2011}]{ruan:1996} 
\'{O} Ruanaidh J. \& Fitzgerald W., 1996, Numerical Bayesian Methods Applied to 
Signal Processing. Springer Verlag:New York
\bibitem[\protect\citeauthoryear{P\'{a}l}{2011}]{pal:2011} P\'{a}l, A., 2011, 
MNRAS, 420, 1630
\bibitem[\protect\citeauthoryear{Podsiadlowski et al.}{2010}]{podsiadlowski:2010}
Podsiadlowski, P., Rappaport, S., Fregeau, J. M. \& Mardling, R. A., 2010, ApJ, 
submitted (astro-ph:1007.1418)
\bibitem[\protect\citeauthoryear{Porter \& Grundy}{2011}]{porter:2011} 
Porter, S. B. \& Grundy, W. M., 2011, ApJ, 736, 14
\bibitem[\protect\citeauthoryear{Ragozzine \& Holman}{2010}]{ragozzine:2010} 
Ragozzine, D. \& Holman, M. J., 2010, astro-ph:1006.3727
\bibitem[\protect\citeauthoryear{Sanchis-Ojeda \& Winn}{2011}]{sanchis:2011} 
Sanchis-Ojeda, R. \& Winn, J. N., 2011, ApJ, 743, 61
\bibitem[\protect\citeauthoryear{Sartoretti \& Schneider}{1999}]{sartoretti:1999} 
Sartoretti, P. \& Schneider, J., 1999, A\&AS, 14, 550
\bibitem[\protect\citeauthoryear{Sato \& Asada}{2009}]{sato:2009} 
Sato, M. \& Asada, H., 2009, PASJ, 61, 29
\bibitem[\protect\citeauthoryear{Seager \& Mall\'{e}n-Ornelas}{2003}]{seager:2003} 
Seager, S., \& Mall\'{e}n-Ornelas, G., 2003, ApJ, 585, 1038
\bibitem[\protect\citeauthoryear{Simon et al.}{2009}]{simon:2009} 
Simon, A. E., Szab\'{o}, Gy. M. \& Szatm\'{a}ry, K., 2009, EM\&P, 105, 385
\bibitem[\protect\citeauthoryear{Skilling}{2004}]{skilling:2004} 
Skilling, J. 2004, in Fischer R., Preuss R., Toussaint U. V., eds,
American Institute of Physics Conference Series Nested Sampling. pp 395–405
\bibitem[\protect\citeauthoryear{Szab\'{o} et al.}{2006}]{szabo:2006} Szab\'{o}, 
Gy. M., Szatm\'{a}ry, K., Div\'{e}ki, Zs. \& Simon, A., 2006, A\&A, 2006, 450, 
395
\bibitem[\protect\citeauthoryear{Taylor}{1992}]{taylor:1992} Taylor, S. R., 
1992, ``Solar system evolution: a new perspective. an inquiry into the chemical 
composition, origin, and evolution of the solar system'', Cambridge University 
Press, Cambridge, UK
\bibitem[\protect\citeauthoryear{Torres et al.}{2004}]{torres:2004} Torres, G. 
Konacki, M., Sasselov, D. D. \& Jha, S., 2004, ApJ, 614, 979
\bibitem[\protect\citeauthoryear{Torres}{2007}]{torres:2007} 
Torres, G. et al., 2007, ApJ, 671, 65
\bibitem[\protect\citeauthoryear{Torres et al.}{2011}]{torres:2011} 
Torres, G. et al., 2011, ApJ, 727, 24
\bibitem[\protect\citeauthoryear{Tusnski \& Valio}{2011}]{tusnski:2011} 
Tusnski, L. R. M. \& Valio, A., 2011, ApJ, 743, 97
\bibitem[\protect\citeauthoryear{Valencia  et al.}{2006}]{valencia:2006} 
Valencia, D., Sasselov, D. D. \& O'Connell, R. J.,  2006, Icarus, 181, 545
\bibitem[\protect\citeauthoryear{Vegetti et al.}{2010}]{vegetti:2010} 
Vegetti, S., Koopmans, L. V. E., Bolton, A., Treu, T. \& Gavazzi, R., 2010,
MNRAS, 408, 1969
\bibitem[\protect\citeauthoryear{Williams et al.}{1997}]{williams:1997} 
Williams, D. M., Kasting, J. F. \& Wade, R. A., 1997, Nature, 385, 234
\bibitem[\protect\citeauthoryear{Williams et al.}{2011}]{williams:2011} 
Williams, D. M., 2011, Oral Presentation, \emph{Kepler Science Conference}, NASA
Ames, California
\end{thebibliography}
\end{document}